\documentclass[12pt]{article}

\usepackage{graphics,epsfig}
\usepackage{amsbsy}
\usepackage{amsfonts}
\usepackage{amsmath}
\usepackage{graphicx}

\def\id{\protect{{1 \kern-.28em {\rm l}}}}
\def\k{\kappa}

\def\p{{\partial}}
\def\nn{\nonumber}

\def\dalemb#1#2{{\vbox{\hrule height .#2pt
        \hbox{\vrule width.#2pt height#1pt \kern#1pt
                \vrule width.#2pt}
        \hrule height.#2pt}}}

\let\a=\alpha \let\b=\beta \let\g=\gamma \let\d=\delta \let\e=\epsilon
\let\z=\zeta  \let\th=\theta  \let\k=\kappa
\let\l=\lambda \let\m=\mu \let\n=\nu \let\x=\xi \let\p=\pi 
\let\s=\sigma \let\t=\tau   \let\c=\chi 
 \let\vep=\varepsilon
\let\w=\omega      \let\G=\Gamma \let\D=\Delta \let\Th=\Theta \let\L=\Lambda
 \let\P=\Pi \let\S=\Sigma  
\let\C=\Chi \let\W=\Omega
\let\la=\label \let\ci=\cite 
  
\def\nn{\nonumber} \def\bd{\begin{document}} \def\ed{\end{document}}
\def\ds{\documentstyle} \let\fr=\frac \let\bl=\bigl \let\br=\bigr
\let\Br=\Bigr \let\Bl=\Bigl
\let\bm=\bibitem
\let\na=\nabla
\def\tU{{\widetilde U}}
\let\pa=\partial \let\ov=\overline
\def\ie{{\it i.e.\ }}
\newcommand{\be}{\begin{equation}}
\newcommand{\ee}{\end{equation}}
\def\ba{\begin{array}}
\def\ea{\end{array}}
\def\ft#1#2{{\textstyle{{\scriptstyle #1}\over {\scriptstyle #2}}}}
\def\fft#1#2{{#1 \over #2}}
\def\F#1#2{{ F_{#1}^{(#2)} }}
\def\cF#1#2{{ {\cal F}_{#1}^{(#2)} }}

\def\R{{\bf R}}
\def\sst#1{{\scriptscriptstyle #1}}
\def\oneone{\rlap 1\mkern4mu{\rm l}}
\def\e7{E_{7(+7)}}
\def\td{\tilde}
\def\wtd{\widetilde}
\def\im{{\rm i}}
\newcommand{\ho}[1]{$\, ^{#1}$}
\newcommand{\hoch}[1]{$\, ^{#1}$}
\newcommand{\bea}{\begin{eqnarray}}
\newcommand{\eea}{\end{eqnarray}}
\newcommand{\ra}{\rightarrow}
\newcommand{\lra}{\longrightarrow}
\newcommand{\Lra}{\Leftrightarrow}
\newcommand{\ap}{\alpha^\prime}
\newcommand{\bp}{\tilde \beta^\prime}
\newcommand{\cB}{{\cal B}}
\newcommand{\cO}{{\cal O}}
\newcommand{\vecx}{\vec{x}}
\newcommand{\vecy}{\vec{y}}
\newcommand{\vecp}{\vec{p}}
\newcommand{\vecq}{\vec{q}}
\newcommand{\tr}{{\rm tr} }
\newcommand{\Tr}{{\rm Tr} }

\newcommand{\cL}{{\cal L}}
\newcommand{\cA}{{\cal A}}
\newcommand{\cD}{{\cal D}}
\def\sst#1{{\scriptscriptstyle #1}}

\def\ve{\varepsilon}
\def\vf{\varphi}
\def\F{\Phi}
\def\wg{\wedge}

\newcommand{\wt}{\widetilde}
\newcommand{\oh}[1]{{\cal O}( #1 )}
\newcommand{\largeoh}[1]{{\cal O}\left( #1 \right)}
\def \foot {\footnote}
\def \bi{\bibitem}

\def \tr {{\rm tr}}
\def \ha {{1 \over 2}}
\def \td {\tilde}
\def \ci{\cite}
\def \N {{\mathcal N}}
\def \ww {\Omega}
\def \const {{\rm const}}
\def \ss {\sum_{i=1}^3 }
\def \t {\tau}
\def\S{{\mathcal S} }
\def \nn {\nu}
\def \XX {{\rm X}}

\def \lra {\leftrightarrow}
\def \vom {{\bar \omega}}
\def \E {{\mathcal  E}} \def \J {{\mathcal  J}}
\def \YY {{\rm Y}}

\def \d {\del}
\def \rJ {{J}}
\def \sms {sigma models\ }
\def \sm {sigma model\ }
\def \L {\Lambda}
\def \gl {\ell}
\def \tr {{\rm tr\ }}
\def\z{\zeta}
\def\zi{\zeta_1}
\def\zii{\zeta_2}
\def\K{\mbox{K}}
\def\eE{\mbox{E}}   \def \vt {\vartheta}
\def \vr {\varrho}
\def \wup {w}

\def\dg{\dagger}
\def\a{\alpha}
\def\b{\beta}
\def\e{\varepsilon}
\def\p{\phi}
\def\ap{\alpha^\prime}
\def\I{{\cal I}}

\def\R{{\bf R}}
\def\Z{{\bf Z}}
\def\C{{\bf C}}
\def\P{{\bf P}}
\def\xb{{\bar X}}
\def\Tr{{\rm  Tr}}
\def\tr{{\rm  tr}}

\def \bs {\bar \s}
\def \btau {\bar \tau}

\def \del{\partial}
\def \a {\alpha}
\def \aa {{\a'}}
\def\g{\gamma}
\def\s{\sigma}
\def\z{\zeta}
\def\zi{\zeta_1}
\def\zii{\zeta_2}
\def\ov{\over}

\def\I{{\cal I}}
\def\J{{\mathcal J}}
\def \ok {{1\ov \k}}
\def\LL{{\mathcal L }}
\def \jL {{J}}
\def \om {\omega}
\def \cL {{\mathcal L}} \def \cH {{\mathcal H}}
\def\E{{\mathcal E}}
\def\w{\omega}
\def\b{\beta}
\def\l{\lambda}
\def\eps{\epsilon}
\def\vep{\varepsilon}
\def \De {{\mathcal D}}

 \def \cV {{\cal V}}
\def  \Jt {  {J}_{\rm tot}    }

\def \k {\kappa}
\def\foot{\footnote}
\def \four{{\textstyle {1\ov 4}}}
 \def \third { \textstyle {1\ov 3
}}
\def\det{\hbox{det}}
\def \ci {\cite}

\def \foot {\footnote}
\def \bi{\bibitem}

\def \tr {{\rm tr}}
\def \ha {{1 \over 2}}
\def \tid {\tilde}
\def \vv {{\rm v}}
\def \tl {{\tilde \l}}
\def \XX {{\rm X}}
\def \ta {{\tilde \a}}
\def \fo { {1\ov 4}}
\def \ep {\epsilon}
\def \inti {{\int^{2\pi}_0 {d \sigma \ov 2 \pi}}}

\def \d {\partial}
\def \K {{\rm S}}
\def \el {\ell}
\def \Tr {{\rm Tr}}
\def \P {\Phi}
\def \l  {\lambda}
\def \tl {{\tilde \l}}
\def \bl {{\tilde \l}}
\def \const {{\rm const}}
\def \V {v}

\def \bv {v^*}
\def \vv {{\rm v}}
\def \LL {{\mathcal L}}
\newcommand{\PV}[1]{P_{\!\!_{V_{#1}}}}

\def \bL {\ell}
\def \M {{\mathcal M}}
\def \N {{\mathcal N}}
\def \S {{\rm S}}
\def \vn {\vec n}
\def \tl {\td \l}
\def \td {\tilde}
\def \Prod {\Pi}
\def \O {{\mathcal O}}
\def \Q {{\rm  Q}}
\def \D {\Delta}
\def \N {{\mathcal N}}
\def\tN{{\tilde N}}

\def \m {\mu}
\def \vs {\vec \s}
\def \ie {i.e.}

\def \cD {{\cal D}}

\def  \le  {\l_{\rm eff}}

\def \rS {{\rm S}}
\def\as{{\a}}
\newcommand{\bra}[1]{\mbox{$\langle #1 |$}}
\newcommand{\ket}[1]{\mbox{$| #1 \rangle$}}

\def\thb{\bar{\theta}}
\def\Thb{\bar{\Theta}}
\def\barp{\bar{p}}
\def\barq{\bar{q}}
\def\barc{\bar{c}}
\def\bard{\bar{d}}
\def\e{\epsilon}

\def \bi{\bibitem}
\def \la {\label}

\def \l {\lambda}
\def\foot{\footnote}
\def \tl  {{\tilde \l}}
\def \sql {{\sqrt \l}}
\def \adss {$AdS_5 \times S^5$\ }
\newcommand{\rf}[1]{(\ref{#1})}
\def \ov {\over}

\def\th{\theta}
\def\Th{\Theta}
\def\vth{\vartheta}

\def\vth{\vartheta}
\def\ra{\rightarrow}
\def\N{{\cal N}}
\def\F{{\cal F}}
\def\cc{\circ}
\def\eqv{\equiv}

\def\ni{\noindent}
\def \ha{{1\ov 2}}
\def \bw {{\rm w}}

\def\r{{\rm r}}

\def \cT {{\cal T}}
\def \no {\nonumber}
\def \J {\mathcal{J}}
\def \del {\partial}

\def \bps {{\bar \psi}}
\def \sqbl {\sqrt{\bar \lambda}}

\def\dF{\dot{F}}
\def\dG{\dot{G}}
\def\df{\dot{f}}
\def \E {{\cal E}}

\def \S {{\cal S}}
\def \J {{\cal J}}

\def\ms{\mathcal{S}}
\def\mj{\mathcal{J}}
\def\soj{\fr{\ms}{\mj}}
\def \R {{\bf R}}
\def \om {\omega}
\def \tH {\widetilde H}
\def \bE {\bar E}
\def \x {{\cal X}}

\def \hV {{\hat V}}

 \def \bb {\bar \beta}
\def \W {{\cal E}}

\def \bi{\bibitem}
\def \la {\label}

\def \l {\lambda}
\def\foot{\footnote}
\def \tl  {{\tilde \l}}
\def \sql {{\sqrt \l}}
\def \sqtl {{\sqrt {\tilde \l}}}

\def \HH {{\rm E}}
\def \cS {{\cal S}}
\def \cL {{\cal L}}
\def \ka {\kappa}

\def \adss {$AdS_5 \times S^5$\ }
\def \arccot {{\rm arccot}}

\def \D {\Delta}
\def \thet {\theta}
 \def \t {\tau}
 \def \p {\phi}
 \def \r {\rho}
 \def \rN {{\rm N}}
 \def\tw{{\tilde w}}
 \def\hJ{{J}}
 \def\hw {{w}}
 \def\hl{{\lambda}}
 \def\hth{{\theta}}
 \def\NN{{\cal N}}
 \def \bv {{ \bar w}}
\def \vn {{\vec n}}
\newcommand{\sfrac}[2]{{\textstyle\frac{#1}{#2}}}
\def \bl {{ \bar \lambda}}
\def \bp {{\bar p}}
\def \bu {{\bar u}}
\def \sha {\sfrac{1}{2}}
\def \w {\omega}
\def \ov {\over}
\def \vl { \vec \ell}
\def \varpi {{\rm w}}
\def \OO {{\cal O}}
\def \bG {\bar \G}

\def \c {\gamma}

\def \ss {{\rm s}}
\def \rE {{\rm E}}

\def \rK {{\rm K}}
\def \ve {\varepsilon}
\def \pa{\partial}
\def \I {{\cal I}}
\def \LL {{\cal L}}
\def \ep {\epsilon}
\def \R {{\rm R}}
\def \tilt {{\tilde t}}

\def\pic #1#2{\hbox{\lower#1pt\hbox{~\mbox{\epsfxsize=20truemm \epsffile{#2}}}}}

\def\pic #1#2#3{\hbox{\lower#1pt\hbox{~\mbox{\includegraphics[scale=#3]{#2}}}}}

\def \bt {\bar\theta}
\def \te {\theta}

\def \cc {{\rm f}}
\def \d {\delta}

\def \cL {{\cal L}}
\def \S  {{\cal S}}
\def \pp {{q}}
\def \vt {\vartheta}
\def \mm {{\cal  \ell}}
\def \Z {{\cal Z}}
\def \pa {\partial}
\def \C {{\cal C}}
\def \be {\bea}
\def \ee {\eea}
\def \c {\gamma}  \def \d {\delta}
\def \eps {\epsilon}

\def \bp {\begin{pmatrix}}  \def \ep {\end{pmatrix}}

 \def \T {{\cal T}}

\def \te {\theta}

\def \r {\rho}
\def \bp {\begin{pmatrix}}  \def \epm {\end{pmatrix}}
\def \ha {{\textstyle{1 \ov 2}}}
\def \om {\omega}
\def \r {\rho}
\def \S {{\cal S}}
\def \z {\chi}

\def \sn {{\rm sn}}
\def \dn {{\rm dn}}

\def \rr {\r_-}

\begin{document}
\overfullrule=0pt
\parskip=2pt
\parindent=12pt
\headheight=0in \headsep=0in \topmargin=0in \oddsidemargin=0in

\vspace{ -3cm} \thispagestyle{empty} \vspace{-1cm}

\vspace{ -3cm} \thispagestyle{empty} \vspace{-1cm}
\begin{flushright} Imperial-TP-AT-2009-5
\end{flushright}
\begin{center}
 \vspace{2cm}
{\Large\bf
 Semiclassical rigid strings with two spins in $AdS_5$
 }

 \vspace{.5cm} {
  A. Tirziu$^{a,}$\footnote{atirziu@purdue.edu}
 and A.A.
 Tseytlin$^{b,}$\footnote{Also at
 Lebedev  Institute, Moscow.\ \
  tseytlin@imperial.ac.uk
 }}\\
 \vskip 0.3cm

{\em
$^{a}$Department of Physics, Purdue  University,\\
W. Lafayette, IN 47907-2036, USA.\\
\vskip 0.08cm
$^{b}$  The Blackett Laboratory, Imperial College,
London SW7 2AZ, U.K. }

\end{center}

 \begin{abstract}
 Semiclassical  spinning string  states  in  $AdS_5$ are, in general,
  characterised by  the
 three  $SO(2,4)$   conserved  charges: the energy $ E $
 and the  two spins  $S_1$    and $ S_2$.
 We discuss several examples  of explicit  classical solutions
 for rigid closed strings of (bended)  circular shape with two non-zero spins.
 In particular, we identify a solution  that  should  represent
 a state that has minimal energy for
 large values  of  the two equal   spins. Similarly
  to  the spiky string  in  $AdS_3$,
 in  the large spin limit  this string  develops   long ``arcs''
 that stretch towards  the boundary of   $AdS_5$. This
 allows the string  to increase the spin
 while having the  energy growing  only logarithmically with   $S=S_1 + S_2$.
 The large spin asymptotics of   such  solutions
 is effectively  controlled by their   near-boundary parts
 which, as in the spiky string case,  happen to be   $SO(2,4)$   equivalent to
  segments of the  straight folded spinning string.
 As  a result,  the    coefficient of the $\log S$  term
 in the  string  energy should  be  given, up to an  overall   $3/2$ coefficient,
 by   the same universal  scaling function  (cusp anomaly) as in the
 folded string  case, to all orders in the  inverse string tension
 or strong-coupling expansion.

\end{abstract}
\newpage

\def \eE {{\rm E}}
\def \rE {{\rm E}}
\def \dn {{\rm dn}}
\def \m {\mu}
\def \n  {\nu}
\def \N {{\cal N}}

\renewcommand{\theequation}{1.\arabic{equation}}
 \setcounter{equation}{0}

\setcounter{equation}{0} \setcounter{footnote}{0}
\setcounter{section}{0}

\def \ads {$AdS_5$ \ }

\section{Introduction}

Trying to achieve  a better understanding of the spectrum of strings in \adss
and thus of strong-coupling expansion of the  $\N=4$ SYM anomalous dimensions
it is of interest to
 study generalizations of  folded \ci{gkp,veg}  and spiky  \ci{kru}
strings  with single spin
in $AdS_3$ part of $  AdS_5$  to the case of strings moving in  full $AdS_5$ and
carrying  two spins.
 The  dimension of $AdS_5$ space  implies that generic
states may  be labeled by the  values of the three $SO(2,4)$  Cartan  generators
 $(E, S_1, S_2)$.
Such semiclassical states   should describe
  strong-coupling behaviour of dimensions
of  gauge-theory  operators  outside  the
$SL(2)$ sector represented, e.g., by operators
  like\foot{Here $\Phi^k$  stands  for some combination of
  fields of SYM theory (which may enter at different places under the trace)
  with $k$ being small and fixed. In the semiclassical string limit in $AdS_5$
  that we will consider below  only the values of the spins $S_1$ and $S_2$
   will matter, i.e.  will  be ``visible'' on the string theory side.}
   \ Tr$  [(D_{0} + D_3)^{S_1} (D_{1}  + i D_2)^{S_2}\   \Phi^k]$.

One may expect that   for large spins  $S_1, S_2    \gg 1 \  $   the string
should stretch
towards the boundary  and the  semiclassical
 states   with minimal energy for  given  values of the  spins should then
 again have  the energy   scaling logarithmically with the spins,
$E-S \sim \ln S + ..., \   S= S_1 + S_2 $.
Indeed, as we shall discuss below,  for the particular   circular strings
with $S_1=S_2$ case  one  finds
$E-S = \frac{3}{2} f(\l ) \ln S +...$,
 where $f(\l)= { \sql \ov \pi}  +  ... $ is the same
 scaling function as in the  one-spin  folded string
 case.
 The  coefficient  of the leading  $\ln S$ term  is
   controlled by the asymptotic  large spin limit of the  solution
 which  happens to be  universal. The extra   factor of $3/2$
  is due to 3 ``arcs''  that the  large-spin  circular string
 has, compared to 2 ``arcs'' of the folded string spinning around zero.
 Remarkably,  the same behaviour  of the energy
  was  found very recently
  from the asymptotic Bethe ansatz
  approach at weak coupling   in
    \ci{rej}.

 The large spin limit  corresponds to the case  when  some parts  of the
 string   approach the boundary of \ads;  it
 may thus be of interest  also for constructing new Wilson loop surfaces   for open
 strings ending at the boundary.
As was shown in \ci{krtt}, the  large spin   limit of the
folded   string in $AdS_3$ is related  via  an analytic continuation  and an $SO(2,4)$
 transformation
to the  open string  solution ending on a null cusp at the boundary  \ci{krus,am1,am2}.
This suggests that  asymptotic limits  of more general  solutions in \ads
may be  also  used for  constructing  interesting  open-string  solutions
lying outside $AdS_3$  (cf. \ci{am3,kruu,dr}).

 Starting  with the bosonic string  in conformal gauge in $AdS_5$  space
 \be
 ds^2 = - \cosh^2 \rho \ dt^2 + d\rho^2  + \sinh^2 \rho \ (d \theta^2 + \cos^2 \theta \ d \phi_1^2 +
 \sin^2 \theta \ d \phi_2^2)  \ , \la{io}
 \ee
 a rigid rotating 2-spin  string  ansatz is  ($0 < \s \leq 2\pi$)  \ci{ft2}\foot{
An equivalent form of this ansatz  is:
$t= \k \tau, \ \r=\r(\s), \ \te'= { \pi \ov 4}, \ \phi'_1 = \om_1 \tau + \te(\s) , \
\phi'_2 = \om_2 \tau - \te(\s)$.
This follows from  writing this solution in embedding   coordinates
and applying a global $SO(4)$   rotation.}
 \be
 t = \kappa \tau \ , \ \ \ \ \    \r= \r(\s)\ , \ \ \ \  \te=\te (\s) \ , \ \ \
 \phi_1 = \om_1 \tau \ , \ \ \  \phi_2 = \om_2 \tau \ . \la{jk} \ee
 The  simplest  circular solution of that  type  is a  round  string with
 $\r=\r_0=\const , \ \te= {\pi \ov 4} , \ \ \om_1=\om_2$
 and thus with $S_1=S_2$ found in
  \ci{ft2}.
 It does not, however, represent a state with a
 minimal energy for given values of the spins (and is, indeed, unstable under small fluctuations
 for large enough value of the spin parameter \ci{ft2}).\foot{As was suggested in  \ci{ptt2}, possible
 gauge-theory  duals  of such circular strings are operators  built out of self-dual part of gauge
 field strength
 \ci{fhz}. Like spiky strings   corresponding to ``excited''  states  in the band of states
 in the $sl(2)$ sector \ci{bgk,fkt}, such  circular strings should
 correspond to  higher-level or ``excited'' 2-spin states.}
 To get a stable  lower-energy solution with $S_1=S_2$ one is to relax
 the $\r=\const$ condition,   allowing the string to develop, in the large spin limit,
  long arcs  stretching to infinity (i.e. the boundary of $AdS_5$)
  and carrying most of the energy.

 A general approach to finding  such rigid string solutions  in $S^5$ or $AdS_5$  was
 developed in \ci{afrt} using the reduction of the conformal-gauge string  sigma model
 to the 1-d  Neumann  integrable  model.\foot{A more general rigid string
 ansatz  where  in addition to
 $ \r= \r(\s),  \   \te=\te (\s) $ one  has
 $\phi_1 = \om_1 \tau  + \a_1(\s) , \   \phi_2 = \om_2 \tau   + \a_2(\s) $
 and where the  corresponding 1-d system is the Neumann-Rosochatius one was
 considered in \ci{art}. Since in this case  the string is stretched  not
 only in $\te$ but also in the other two
 angles, one expects that such solutions should have  more energy  for given  values of the spins;
 we will not consider this more general case  in what follows.}
 The solutions in $S^5$ and $AdS_5$ are  closely related via an analytic continuation.
 Starting with the $R^{2,4}$
 embedding coordinates   satisfying   $|Y_{05}|^2-|Y_{12}|^2-|Y_{34}|^2=1$
 ($Y_{nm} \equiv  Y_n + i Y_m$)
\be
Y_{05}= y_0 e^{i t},
 \quad \quad Y_{12}=   y_1  e^{i \phi_1}, \quad \quad
  Y_{34}=   y_2  e^{i \phi_2} \  , \ \ \ \ \ \          y_0^2 - y_1^2 - y_2^2 =1  \ , \la{yq}
  \ee
  where the  choice of coordinates in \rf{io} corresponds to
\be y_0= \cosh \rho\ , \ \ \ \  y_1= \sinh \r \ \cos \te  \ , \  \ \
y_2 = \sinh \r \ \sin \te \ , \la{uuu}
\ee
 and assuming that $y_a=y_a(\s)= y_a(\s + 2 \pi)$  and $t= \k \tau,\  \phi_i= \om_i \tau$
 one finds that the equations for $y_a$ are those of a harmonic oscillator
 constrained to move on a 2d hyperboloid -- an integrable system with 2 integrals of motion
 $b_1,b_2$  with
  $b_1+b_2 = \k^2 + \om^2_1 + \om_2^2$.

 Following the  discussion in \ci{afrt}, the closed string solutions  will
  be  parametrised by  the three ``frequencies'' $\om_a=(\om_0, \om_1,\om_2$), \ $ \om_0\equiv\ka$,
  as well  by the  two integrals of motion $b_i$.   Four of  these
 parameters, say  $(\om_i,b_i)$,
 may be viewed as independent  coordinates on the  moduli space of such  solitons.
The closed string periodicity condition
in $\s$ implies that   solutions will be classified
by  two    integer ``winding numbers''   $n_i$  related to  $\om_a$ and $b_i$.
 In general, the energy
$E$ will be a function not only of $S_1,S_2$ but also of  the values of $n_i$.
Depending on the values of these parameters  the string's   shape may be of two distinct types:
(i)
 {\it ``folded''}, i.e. having  topology of an interval,
 or  (ii)  {\it  ``circular''},  i.e. having  topology of a circle.
 A folded string may be  straight as in the one-spin  case  \ci{gkp}
 or bent.
 A ``circular'' string  may  be  a round circle as in  \ci{ft2}  or
 may  have a more general
 ``bent circle'' shape.
 To have  a folded string  we need all derivatives
  $y'_a$  vanishing at the two  points  of the $\s$ interval.
 To have a bend  we need  only one  out of the two independent coordinates having their
  derivative vanishing
 in a middle point of the $\s$ interval.\foot{If that vanishing  happens in $k$ points
  we will have a string
 with $k$ bends.}

 It is instructive to recall \ci{afrt}   how  classical  solutions   with such shapes
appear  in the flat $R^{1,4}$ Minkowski space
which corresponds to the $\r\to 0$ limit  of \rf{io}   or the limit of
\rf{yq} when $y_0\to 1$ and $y_1,y_2 $ are small.
The 5 independent string coordinates  can thus be parametrised by
$t= \k \tau$ and $Y_{12}$ and $Y_{34}$ in \rf{yq}  and solution of  interest is given by
\be \la{taak}
\om_1= n_1 \ , \ \  \om_2 = n_2 \ , \ \ \
 y_1  = a_1   \sin (n_1 \s) \ ,  \ \ \ \ \ \
y_2  = a_2  \sin [n_2 (\s + \s_0) ]  \ ,    \ee
where  $n_i$ are integers,  $\s_0$=const    and
$  \k^2 =   n^2_1 a_1^2 + n^2_2 a_2^2 $.
 Then  the  energy and  the two spins  are
$
 E= {\k  \ov \a'}  , \ \  S_i = {n_i a^2_i \ov 2 \a'} $, i.e.
$
 E= \sqrt { { 2 \ov \a'} ( n_1 S_1 + n_2 S_2 ) } $.
To get  the states on the leading  Regge trajectory
(having minimal energy for given  values of the { two}  spins) one  is to choose
 $n_1=n_2=1$.\foot{One is also to set  $\s_0= {\pi \ov 2}$
 since otherwise the $O(4)$ angular momentum has  other  nonzero components
 apart from $S_{12}=S_1, \ S_{34} = S_2$ and thus the solution can be  rotated  to a
single spin one. For example, for $\s_0=0$ the solution  is equivalent to the straight folded  string.}
 The shape of the string depends on the values of $\s_0$ and $n_1,n_2$:
 it can be either  circular or folded.\foot{If $\s_0\ov \pi $ is irrational then the string
 always has a ``circular''
 shape and,  in general, will not be lying in one plane, i.e. will have one or several bends \ci{afrt}.
For rational values of $\s_0$ the string can be
either circular or folded,  depending  on the values of $n_1,n_2$.
For $\s_0=0$ if  both $n_1$ and $n_2$ are either even or odd and different
then the string is folded and has several bends
(in the 13 and 24 planes).
If  ${\s_0 \ov \pi} ={1\ov 2 n_2}$ and   $n_1=n_2$ the string is an ellipsoid,
becoming a round circle in the special case of $a_1=a_2$.
The string is also circular if $n_1$ is even and $n_2$ is odd.
If, however, $n_1$ is odd and $n_2$ is even the string is folded and bent.}

The structure of the soliton strings
in   curved  $R_t \times S^5$ or $AdS_5$ case is analogous \ci{afrt}.
Indeed, the equations of motion of the Neumann system are linearized
on the Jacobian of the hyperelliptic curve.
The general  solution for $y_a(\s)$ in \rf{uuu}
   is then  expressed in terms of
 hyperelliptic functions (theta-functions defined on the
  Jacobian of the hyperelliptic genus 2
 Riemann surface).\foot{
 The image of the string in the Jacobian
 (Liouville torus)  winds around two
non-trivial cycles with the winding numbers $n_1$ and $n_2$.
 The size and the shape of the Liouville torus
are governed by the moduli $(\om_i,b_i)$. For given
$n_1,n_2$,  two of the 4 independent  parameters  $(\om_i,b_i)$
are then  uniquely determined by
the  periodicity conditions.}
 The  shape of the physical string
 at  fixed moment of time  lying on the 2d hyperboloid described by $y_a$  in \rf{yq}
  will depend
on the  values of  $n_1,n_2$ and other  moduli parameters
and may be of the bent   folded type or of the circular type.
\foot{The simplest  round-circle  string solution \cite{ft2} mentioned above
 corresponds
to the case $\om_1 = \om_2$ and $b_1 = b_2$.}


  There are few   special cases  when  solutions simplify, i.e. when
   the hyperelliptic surface degenerates into
  an elliptic one  so that  $y_a(\s)$
  can be  expressed  in terms of
  the standard elliptic  functions (as in the one-spin case \ci{veg,gkp,ft1}). Such special solutions
 are    much easier to analyse and potentially  compare to the corresponding
  states on the gauge-theory side.
  These are  the cases
   when  two of the   3 frequencies $(\k=\om_0,\om_1,\om_2)$  or    two
  of the  integrals of motion  $b_1,b_2$ are equal.
 In sections 2 and 4 below we shall consider    two of such special cases:
\be  (i) \  \om_1=\om_2    \ , \ \ \ \ \ \ \ \ \ \ \   (ii)\  \k= \om_2   \ .  \la{eee} \ee
 As we shall see,  in these   cases  the string is  of circular type.
 In the first case  $S_1=S_2$   while in the  second one  $S_1 \not=S_2$.\foot{
 While  the solution corresponding to (i)
 appears to  have minimal energy for given $S_1=S_2$
 this is  likely  not to be the  case  for the solution in (ii): there should be a folded
 bended string solution  that carries less energy   for given $S_1\not=S_2$.
  When discussing  minimal energy for given spins we assume we also choose
  minimal possible values for the ``winding numbers''  $n_1,n_2$.}
The case of $b_1=b_2$ will be discussed  in Appendix B.

 The  solution corresponding  to the  first case  in \rf{eee} was  found in the $S^5$  setting
 in sect.4.2 of \ci{afrt}; its direct $AdS_5$ counterpart  which has $S_1=S_2$
 was  implicit in  sect. 6 of \ci{afrt} and
 was described explicitly in \ci{ kl} (this solution was  also  generalised to a
 non-zero value of one angular momentum in $S^5$ in \ci{r}).

In the first case  with  $ \om_1=\om_2 \equiv \om$  the string sigma model equations
corresponding to \rf{io} can be readily solved by integrating  the equation for  $\te$.
The  integral of the equation for $\r$ is
 given by the conformal  gauge condition, so that  we end up
with
\bea  && \theta'= \frac{c}{\sinh^2 \rho}  \ ,  \label{sqr} \\
&&\rho'^2=\kappa^2 \cosh^2 \rho - \frac{c^2}{\sinh^2 \rho}-\omega^2 \sinh^2 \rho  \ , \la{con}
\eea
 where $c$ is an  integration constant.
   When  $c=0$
 the solution reduces to the single-spin folded string one.
 Since $\te'$   does not vanish for $c\not=0$  (unless at the points where $\r\to \infty$
 which correspond to the large spin  asymptotics)
 this solution  has a  circular   shape.
 The  simplest among such  2-spin solutions  \ci{ft2}
 has ($\sinh^2 \r_0= {c\ov  m} $)
 \be \r(\s)=\r_0=\const\  , \ \ \ \ \ \ \ \    \te(\s) =  m \s\ , \ \ \ \ \ \ \
 m=1,2,3,... \la{cii} \ee
 and describes  a rigid circular string wrapped $m$ times  in $\te$  and  rotating
 in two  planes with equal spins
 $S_1=S_2=S/2$.
 The   energy of  this solution  for large $\S= { S\ov \sql}\gg 1 $ scales is\
 \be \la{faa}
 E-  S = \sql \ [ {3 \ov 4} (2m^2 \S)^{1/3} + ... ]
 \ee
 i.e.  it grows  faster than $\ln S$.
 This solution is  unstable for large enough spin \ci{ft2,ptt2}, suggesting that there  should be
 a similar $S_1=S_2$ solution  having  lower energy for given spins.

 To find  such a  lower energy  2-spin state
 one is  to consider  solutions  of  \rf{con}  with non-constant $\r$:  that will
 allow one to increase  the spin  by stretching parts of the circular-shaped
 string towards the boundary. This is  energetically more  favorable   than
 putting the whole  round  string at large value of $\r$
   as in  the case of   the  ``round circle''   solution \rf{cii}.
In contrast to  spikes \ci{kru}, these stretched arcs will still  have  regular
shape:  the
  induced metric  here $ds^2 = (\r'^2 + { c^2 \ov \sinh^2 \r}) ( - d \tau^2 + d \s^2) $
is everywhere smooth as long as $c\not=0$.

\

 We shall review and clarify
  the corresponding solution \ci{afrt,kl} of eqs. \rf{sqr},\rf{con}  in section 2.
 As we shall discuss in section 3,
 its   large-spin limit  when its $E-S$
  scales as $\ln S$ is effectively  controlled by the asymptotic ``single-arc''
  open-string
 solution corresponding to the case when  $\k=\om_1=\om_2 $.
 This  solution is found to be   equivalent, by an $SO(2,4)$ transformation,
   to the  asymptotic limit of the folded  or spiky string. This
   implies  that  the coefficient of the leading   $\ln S$ term
   should be proportional to the universal scaling  (cusp  anomaly) function;
   that should be true
   to  all orders in the string $\a' = {1 \ov \sql}$ expansion.

   In section 2 we shall   also compute  the first subleading
    coefficient in the large spin expansion in the classical string energy  and
    compare it to the one in the spiky string case  \ci{kru,bftt}.
   Our result for the leading terms in  large-spin expansion
   of the classical energy  of a circle-shaped
    string with two equal spins ($S_1=S_2= \ha S$), winding number $m$
   and $n > {m \ov 2} $ arcs  is
      \begin{equation}
{E}-{S}= \frac{n\sql }{2 \pi} \Big(\ln \frac{16 \pi {S}}{n\sql} -1 +  2
 \ln \sin \frac{\pi m }{n} \Big)  + \mathcal{O}(\frac{1}{{S}})   \ .   \la{sjp}
\end{equation}
The $\ln S$  large spin  asymptotics of the energy of this $S_1=S_2$
solution was first observed  in \cite{kl}.
 The minimal energy for given spins  is found for  $m=1, \ n=3$  when
 \begin{equation}
{E}_{\rm min} - 2{S}_1= \frac{3}{2} \times { \sql \ov  \pi} \Big(\ln {8 \pi {S}_1 \ov \sql} -1 \Big)
+ \mathcal{O}(\frac{1}{{S}}_1) \ .  \la{sjkp}
\end{equation}
 This is  $3 \ov 2$  times
   the expression for the  folded   string   with a single  spin $S_1= \ha S$
   which represents the minimal energy state for given spin in $AdS_3$
   (or the  ground state in the $sl(2)$ sector).

  Remarkably, this   matches  the strong-coupling prediction  following from the very recent
   analysis \ci{rej} of the full  Asymptotic Bethe Ansatz
   equations \ci{bes}.\foot{We thank A. Rej for
  informing us about the results of \ci{rej}.}
This  agreement is, of course, not unexpected as  the scaling (``thermodynamic'') limit
of the full version of the  strong-coupling limit of the Bethe ansatz
equations \ci{afs}     should  reproduce   finite-gap solutions
of the classical  string sigma model \ci{gap,bgk}. Still, the  precise identification
of a particular string solution  that has a   clear space-time interpretation with
 a  particular Bethe root  distribution is,  in general,   non-trivial,
 especially for states   outside simplest rank-one sectors.

In \cite{rej} a $1$-cut Bethe root distribution was found, which allowed
the authors to
compute the leading and subleading\footnote{To get the subleading correction in ABA
one splits the root distribution $\rho(u)=\rho_0(u)+ r(u)$, where $\rho_0$ is the
root distribution with a $1$-cut support. To get the correction $r(u)$, one solves the
resulting integral equation on the whole real axis. Thus, the subleading large $S$
 result
contains information from outside the actual $1$-cut region. On the string side,  this
corresponds to the fact that to get the subleading in large $S$ correction, one needs to use
an extra
information about the  exact solution, not only its leading asymptotic form.}
terms in the large $S$ expansion of the energy. To describe
the strong-coupling solution for a finite value of semiclassical spin
  one should go
beyond the $1$-cut solution of the  ABA equations,  i.e. one  should
identify the $2$-cut
distributions with  elliptic (genus 1)  string solutions
and $3$-cut distributions with solutions
associated to   hyperelliptic (genus $2$) Riemann surface.
The large $S$ asymptotics correspond  to the case when cuts collide.
The rigid $2$-spin solutions that we discuss here
 are generically hyperelliptic;
it should be possible to  determine
 which  root density that solves integral equations following from the
  ABA   should correspond to the
generic hyperelliptic rigid-string  solution with finite (semiclassical)  value of $S$.

 In section 4 we shall consider  the second
 special case --  that of  $\k= \om_2$  in    \rf{eee}.
 This case is closely related to the previous one via an analytic continuation
 in which the roles  of $\k$ and $\om_1$ and $y_0$ and $y_1$ are interchanged
 (as implied by the general discussion in \ci{afrt}).
Here in general $S_1\not=S_2$, but to allow for the existence of a large spin limit
 one is to  go back to the case of $S_1=S_2$.
 The corresponding asymptotic solution is again the  one of section 3.

In section 5 we shall comment on  large spin behaviour of more general solutions  described
by the ansatz \rf{jk} and   make  some concluding remarks.

In Appendix A  we shall  discuss a special case of the  solution  of section 4.
In Appendix B we shall  review  the approach of \ci{afrt}  to solution of equations
corresponding to the rigid string ansatz \rf{jk} and  consider in detail the special
 case of $b_1=b_2$
when the circular string solution is again expressed in terms of elliptic functions.
In this case the  energy  is found to scale  with the large total spin $S=S_1 + S_2
$ as in the round-circle $S_1=S_2$  case, i.e.   $E-S \sim S^{1/3}$.

 \renewcommand{\theequation}{2.\arabic{equation}}
 \setcounter{equation}{0}

\setcounter{equation}{0}

\section{Rigid ``circular''  $S_1=S_2$  solution: $\omega_1=\omega_2$}

Setting $x\equiv y_0=\cosh \r$  in \rf{con} we obtain the equation for $x(\sigma)$
\be
x'^2 = \k^2 x^2 (x^2-1) - c^2  - \om^2 (x^2-1)^2\ ,
\ \ \ \ \  \ \  x\equiv \cosh \r  \ ,
\ ,
\label{yyy}
\ee
or, equivalently  \ci{afrt,kl}
\begin{equation}
x'^2= (\om ^2-\kappa^2) (x^2-a_{-})  (a_{+}-x^2) \ ,           \label{xoe}
\end{equation}
where
\begin{equation}
a_{\pm}= \frac{2 \omega^2 - \kappa^2 \pm \sqrt{\kappa^4-4 c^2 (\omega^2-\kappa^2)}}{2 (\omega^2- \kappa^2)}
\end{equation}
Thus
\begin{equation}
c^2=(a_{+}-1)(a_{-}-1) (\omega^2 -\kappa^2), \quad \quad
\kappa^2= \omega^2 \frac{a_{+}+a_{-}-2}{a_{+}+a_{-}-1}= \omega^2 \frac{\m(2-\n)}{\n+\m-\n\m}  \label{opii}
\end{equation}
where we introduced the parameters $\m$ and $\n$  related to $a_\pm$ by ($0<\m<\n \leq 1$)
\foot{Our notation
are related to those of \ci{kl,r}  by $\mu \to m, \ \nu \to n, \ m \to M, \ n \to N$.}
\begin{equation}
\m=\frac{a_{+}-a_{-}}{a_{+}} , \quad
\quad \n=\frac{a_{+}-a_{-}}{a_{+}-1} , \qquad  a_{+}=\frac{\n}{\n-\m}, \quad \quad a_{-}
=\frac{\n(1-\m)}{\n-\m} \la{mmm}
\end{equation}
$x(\s) $  takes values in
$\sqrt{a_{-}} \leq x \leq \sqrt{a_{+}} $,
i.e. the radial string coordinate  $\rho$
changes in the interval
\be
 \r_- \leq \r \leq \r_+ \ , \ \ \ \ \ \ \ \ a_\pm = \cosh^2 \rho_{\pm}
 \la{rar} \ . \ee
  Since  $x=\cosh \rho \geq 1$ we
   have  $\omega \geq  \kappa$ and
    $1 \leq a_{-} < a_{+}$.
Solution of equation for $\rho$ (\ref{xoe}) is
\begin{equation}
x=\cosh \rho= \frac{\sqrt{a_{-}}}{\dn[c \sqrt{\frac{a_{+}}{(a_{+}-1)(a_{-}-1)}}\sigma,\m]} \ ,  \label{sol1}
\end{equation}
where $\dn$ is the Jacobi  elliptic function.

We will assume
that  $\rho$  starts at its minimum $\rho_{-}$ at $\sigma=0$ and goes to its  maximum
 $\rho_{+}$ at $\sigma=\frac{\pi}{n}$ where $n$ is an integer number.
  To get a closed   string   defined on  $0 \leq \sigma \leq 2\pi$ we need to glue together
   $2 n$ such segments (or  $n$  string ``arcs''   where  $\r$ first
   grows to maximum and then comes back)
    imposing the  periodicity condition
   $\rho(\sigma+2 \pi)=\rho(\sigma)$.
    Since the period of the  $\dn[z,\m]$ function is $2 \rK[\m]$  \foot{Here
    $\rK[\m]=\int_0^{\frac{\pi}{2}} \frac{d \alpha}{\sqrt{1-\m \sin^2 \alpha}}$.
    Below we shall also use   $\rE[\m]=\int_0^{\frac{\pi}{2}} d \alpha \sqrt{1-\m \sin^2 \alpha}$.}
     we have
\begin{equation}
c \sqrt{\frac{a_{+}}{(a_{+}-1)(a_{-}-1)}} \frac{2 \pi}{n}=2 \rK[\m] \ ,  \label{opi}
\end{equation}
i.e.  the periodic solution can be written as
\begin{equation}
x=\cosh \rho= \frac{\sqrt{a_{-}}}{\dn[\frac{\rK[\m]n}{\pi}\sigma,\m]}  \label{solu1}
\end{equation}
At $\sigma=\frac{\pi}{n}$ we indeed have $x=\sqrt{a_{+}}$ since $\dn[\rK[\m],\m]=\sqrt{1-\m}$.
Note  that  (\ref{opi}),(\ref{opii}) imply
\begin{equation}
\omega= \sqrt{\frac{a_{+}+a_{-}-1}{a_{+}}}\frac{\rK[\m]n}{\pi}=\sqrt{\frac{\n+\m-\n
\m}{\n}}\frac{\rK[\m]n}{\pi}
 \label{onl}
\end{equation}
To find  $\theta$ it is more convenient to use the equation for $x(\theta)= x(\te(\s)) $
which follows from  \rf{sqr},\rf{con}
\begin{equation}
\frac{d x}{d \theta}= \pm \frac{ (x^2-1) \sqrt{(a_{+}-x^2)(x^2-a_{-})} }{\sqrt{(a_{+}-1)(a_{-}-1)}}  \label{eft}
\end{equation}
Solving this equation for $\te(x)$  with the initial condition  $\te (x_{min}=\sqrt{a_{-}})=0$
(i.e. $\te(\s=0) =0$)
 we obtain \foot{Here $\Pi[\n,z,\m]=\int_0^{z} \frac{d \alpha}{(1- n \sin^2
 \alpha)\sqrt{1-\m \sin^2 \alpha}}$,\ \
\   $\Pi[\n,\m]\equiv \Pi[\n,\frac{\pi}{2},\m]$.}
\begin{equation}
\theta(x)= \sqrt{\frac{a_{-}-1}{a_{+}(a_{+}-1)}}\bigg(\Pi[\n,\m]-\Pi[\n,\arcsin
 \sqrt{\frac{a_{+}-x^2}{a_{+}-a_{-}}},\m]\bigg) \label{sol2}
\end{equation}
Then
\begin{equation}
\theta(\sqrt{a_{+}})=\sqrt{\frac{a_{-}-1}{a_{+}(a_{+}-1)}}\Pi[\n,\m]  \label{ddi}
\end{equation}
The expression  \rf{sol2}
 is valid for one  half-arc   of the string with  $ \r_- < \r < \r_+$, i.e.
  it gives $\te(\s)= \te (\cosh \r(\s))$  for   $0 \leq \sigma \leq \frac{\pi}{n}$.
Full solution for $\theta(\sigma)$ can be easily obtained using (\ref{sol1}).
 To cover the  $(0,2 \pi)$ \  $\sigma$-interval we  should glue together $2 n$  segments
 given by  \rf{sol2}.
The condition for having a closed string gives
\begin{equation}
\theta(2 \pi)= 2 \pi m = 2 n \theta(\frac{\pi}{n}) \ , \ \ \ \ \ \ \   m=1,2,3,...\ ,  \label{shu}
\end{equation}
where we introduced  an arbitrary  winding number $m$.
Plugging this into (\ref{sol2})  gives
\begin{equation}
\pi \frac{m}{n}=\sqrt{\frac{a_{-}-1}{a_{+}(a_{+}-1)}} \Pi[\n,\m] =
\sqrt{\frac{(1-\n)(\n-\m)}{\n}} \Pi[\n,\m]  \label{cond1}
\end{equation}
We thus  need $m \neq 0$ in order to satisfy this condition.
 One can show that the right hand side in (\ref{cond1}) is
 always smaller than $\frac{\pi}{2}$. This implies the condition on the parameters \ci{kl}
 \be   2 m < n \ , \ \ \ \ {\rm i.e.} \ \ \ \   n \geq 3        \ .  \ee
The minimal choice  is thus $m=1, \ n=3$.

\subsection{Energy and spins}

The  energy  and the two spins  are defined by  ($E= \sql \E, \ \ S_i = \sql \S_i$)
\be  \E= \k \int^{2 \pi}_0  {d\s\ov 2 \pi}   \cosh^2 \r , \ \
\S_1 =  \om_1  \int^{2 \pi}_0    {d\s\ov 2 \pi}  \sinh^2 \r \  \cos^2 \te   , \ \
\S_1 =  \om_2  \int^{2 \pi}_0   {d\s\ov 2 \pi}  \sinh^2 \r \  \sin^2 \te  \la{enk}
\ee
For $\om_1=\om_2$,   periodic $\r(\s)$  and $\te(\s + 2 \pi) = \te(\s) + 2 \pi m$
one can argue that  $\S_1=\S_2$.
  This  follows, e.g.,
from the vanishing  of the integral
$ \int d \te  \  \sinh^4 \r \ \cos 2 \te $    obtained by  converting
  the integral over $\s$ into the integral
over $\te$ using \rf{sqr}.
One can similarly show that
other ``non-Cartan''
  components of the $SO(2,4)$   angular momentum tensor
  vanish (cf. \ci{afrt}), i.e.  the corresponding semiclassical state
  has  only $E,S_1,S_2$  as its  global ``quantum numbers''.

The energy   and the total spin $S=S_1+S_2= 2 S_1 =  \sql \S $ can be written as
\begin{equation}
\mathcal{E}=\frac{n}{\pi} \frac{\eE[\m]}{\n-\m}\sqrt{\n \m (2-\n)}, \quad \quad \mathcal{S}=\frac{n}{\pi}\bigg(
\frac{\n \eE[\m]}{\n-\m}-\rK[\m]\bigg)\sqrt{\frac{ \n+\m -\n \m}{\n}} \ ,  \label{wind}
\end{equation}
where $\eE$  is the standard elliptic function.
Then  for given $n,m$  and $\S$   the energy takes the form
\begin{equation}
\E=\ n \ {\bar \E} (\frac{\S}{n},\frac{m}{n})   \ . \la{er}
\end{equation}
We have checked numerically that for
the minimal choice  $m=1$, $n=3$  the
 equation \rf{cond1}, and the second equation in (\ref{wind})
 admit solution for $\n,\m$ for various values of the spin $\mathcal{S}$.

   In Figure 1 we  present   the profile  of the string in $(\rho, \theta)$ polar coordinates.
\begin{figure}
\epsfig{file=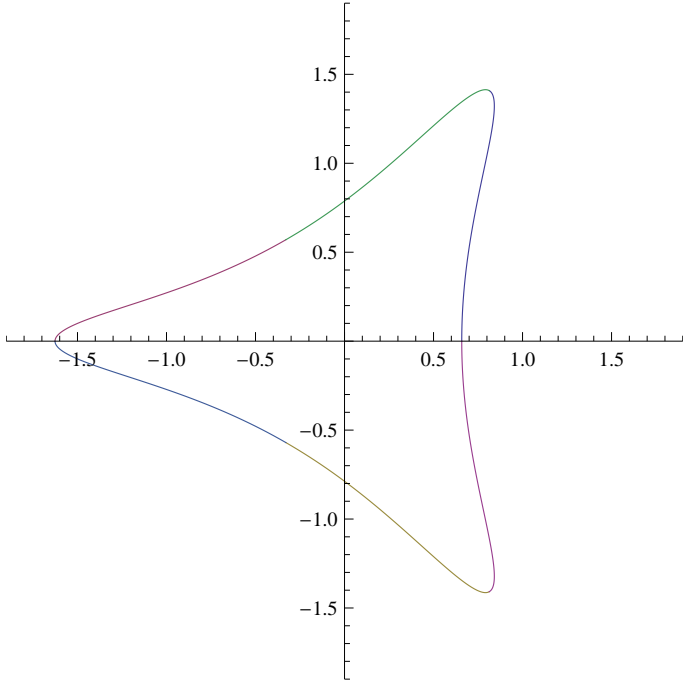, width=9cm}
\caption{Shape of  $S_1=S_2$ string with  $m=1$, $n=3$ and  $a_{-}=1.5$, $a_{+}=6.97$.
}
\label{fig1}
\end{figure}
Let us stress  that  the shape of the string is smooth (no spikes at maxima or minima of $\rho$)
 since  $d \r \over d \theta$ is continuous. This case  is thus ``intermediate''
 between   the  one  of the 1-spin  spiky string\footnote{A similar effect appears also
 in the spiky string when a $S^5$ spin $J$ is included \cite{iktt}.} \ci{kru} and the one
 of  the 2-spin  circular  string in \ci{ft2}.

Let us  now look at some special cases.

\subsection{ ``Round circle'' limit: \ \   $\rho_+ \approx \rho_-$}

The limit when the   variation of the radial distance  $\rho$  with $\s$
(or, equivalently, with  $\theta$)   is small,
i.e. the shape of the string is close to   a  round circle,
may be obtained by taking  $a_+ \to a_-$, i.e.  $ \m \ll 1$  (cf. \rf{mmm}).
Using the expansion of the elliptic integral $\Pi[\n,\m]$
 for small $\m$
\begin{equation}
\Pi[\n,\m]=\frac{\pi}{2 \sqrt{1-\n}}+O(\m)
\end{equation}
the condition (\ref{cond1}) becomes
\begin{equation}
\frac{2 m}{n}\approx \sqrt{1- \frac{\m}{\n}}  \label{stg}
\end{equation}
Since $\n,\m>0$ we  again conclude  that  $2 m<n$.
The  condition \rf{stg} can be  satisfied for generic integer
 $n,m$  if
 $\n$ is also small, i.e. $\n \approx \m \ll 1$.
In this limit
we obtain
\begin{equation}
a_{+}= \cosh^2 \r_+ \ \approx \big(\frac{n}{2m}\big)^2, \quad \quad a_{-}=\cosh^2 \r_-\
\approx \big(\frac{n}{2m}\big)^2(1-\m)
\end{equation}
The  shape of the string is close to a round circle  located at a distance $\r_+ \approx \r_-$
from  the center of $AdS_5$  (its length  may be   approximated by $ 2 \pi m  \sinh^2 \r_+$).
 The energy and total spin are found to be
\begin{equation}
\mathcal{E}=\frac{n^2}{4 \sqrt{2} m^2}\sqrt{n^2-4 m^2}\ ,\qquad \quad \mathcal{S}=\frac{n^2-4 m^2}{4
\sqrt{2}m^2}\sqrt{n^2- 2 m^2}
\end{equation}
The case of  the circular string (where $\r_+=\r_-$)    corresponds to
 fixing, e.g., $m=1$ and taking $n$ large to recover the round  shape of the string.
Then solving for $n$   in terms of $\S$ we find  the same large spin  relation
 as for the circular string of \ci{ft2}:
\be
\E = \S +   { 3 \ov 4}  (2\S)^{1/3} + ...   \  . \la{fg} \ee

\subsection{Large spin  limit}

Let us now consider the  limit  when  the variation   of $\rho$ with $\theta$
 is large
 and  the total   spin  is   large.  This corresponds to   $\n \approx \m \approx 1$ with $\m < \n$.
In this limit
\begin{equation}
\rK[\m] \approx \frac{1}{2}\ln \frac{16}{1-\m} \gg 1  \ .
\end{equation}
Then we can approximate the elliptic integral on the right hand side of  (\ref{cond1})  as
\begin{equation}\la{ooo}
\Pi[\n,\m] \approx \sqrt{\frac{\n}{(1-\n)(\n-\m)}}\bigg(\frac{\pi}{2}- {\rm arcsin}
 \sqrt{\frac{1-\n}{1-\m}}\bigg)
\end{equation}
and eq.(\ref{cond1}) becomes
\begin{equation}
\cos \frac{\pi m}{n} \approx \sqrt{\frac{1-\n}{1-\m}}  \label{lpp}
\end{equation}
which implies  $2 m<n$.
In this limit become
\begin{equation}
\kappa^2= \omega^2 \frac{\m(2-\n)}{\n+\m-\n\m} \approx \omega^2,
\quad \quad \omega \approx \rK[\m] \frac{n}{\pi} \gg 1  \,
\end{equation}
where we used (\ref{onl}). The parameters $\kappa$  and  $\omega$ are
 thus  approximately equal and
large in this limit.
At leading order in large $a_{+}$ the equation (\ref{lpp}) implies
\begin{equation}
a_{-} \approx  \frac{1}{\sin^2 \frac{\pi m}{n}}  \label{ahg}
\end{equation}
Then for  $\mathcal{E}-\mathcal{S}$ we obtain at the leading order
\begin{equation}
\mathcal{E}-\mathcal{S}=\frac{n}{2 \pi}\ln \frac{16}{1-\m}  + ... \label{qlp}
\end{equation}
The total spin $\S= 2 \S_1 = 2 \S_2 $  can be written  as
\begin{equation}
\mathcal{S}=\frac{n}{\pi} \frac{1}{(1-\m) \sin^2 \frac{\pi m}{n}}-\frac{n}{4 \pi \sin^2 \frac{\pi m}{n}}
\Big[2 +  \cos \frac{2\pi m}{n}(1+\ln \frac{1-\m}{16})\Big]+O(1-\m)  \label{spj}
\end{equation}
 and is   large  since   $\m \rightarrow 1$.
Solving  (\ref{spj}) for $1-\m$ in term of $\mathcal{S}\gg 1$  and plugging  it into
the energy in \rf{wind}
we find
\begin{equation}
\mathcal{E}-\mathcal{S}= \frac{n}{2 \pi} \Big[\ln \frac{16 \pi \mathcal{S}}{n} -1 +  2
 \ln \sin \frac{\pi m}{n} \Big]  + \mathcal{O}(\frac{1}{\mathcal{S}})     \la{sj}
\end{equation}
 The minimal energy solution is found for  $m=1$, $n=3$  when we get\footnote{To get the $ \mathcal{O}(\frac{1}{\mathcal{S}_1})$ term we used the next order expansion

 $\Pi[\n,\m] \approx \sqrt{\frac{\n}{(1-\n)(\n-\m)}}\bigg(\frac{\pi}{2}- {\rm arcsin}
 \sqrt{\frac{1-\n}{1-\m}}\bigg)+[\frac{1}{2}-\frac{\sqrt{\n}}{4}]\ln \frac{16}{1-\m}+\frac{\sqrt{\n}}{4}+\mathcal{O}(\m-1)$.} ($\S=2 \S_1$)
 \begin{equation}
\mathcal{E}- 2\mathcal{S}_1= \frac{3}{2 \pi} \Big[\ln ({8 \pi \mathcal{S}}_1) -1 \Big]
+ \frac{9}{8 \pi^2 \mathcal{S}_1}\Big[ \ln (8 \pi  \mathcal{S}_1) +{1 \ov 2}  \Big]+\mathcal{O}(\frac{1}{\mathcal{S}_1^2}) \ .  \la{sjk}
\end{equation}
 Remarkably, the factor  multiplying $3 \ov 2$  in the first term of the right hand side in (\ref{sjk})
  is exactly the same  as for the  folded   string   with
  a single  spin $\S_1= \ha \S$.
  This conclusion  is precisely the one  following (in the strong-coupling limit)
  from the analysis  of the Asymptotic Bethe Ansatz  equations \ci{rej}.
Let us also remark that the coefficient of the $\frac{\ln \mathcal{S}_1}{\mathcal{S}_1}$ term is again as in the  folded
case; more precisely, it is $\frac{1}{2}$ times the coefficient of the $\ln \mathcal{S}_1$ squared.\footnote{As in folded string case  this is the
 pattern    implied by the functional
relation \cite{bftt}, i.e.

 $E-S = f \ln (S + \frac{f}{2}  \ln S + ...)  + ...= f \ln S + \frac{f^2}{2} \frac{\ln S}{S}+...$,   where now  $f=\frac{3}{2 \pi}$. }

For general $n$ (and $m=1$)  the   expression  \rf{sj} may be compared
 to  the one found for the
 single-spin string  with $n$ spikes  \ci{kru,bftt,fkt}\footnote{See also \cite{dorey} for a derivation using the spectral curve approach.}:
 \be
 \mathcal{E}_{\rm spiky} - {\S_1}= \frac{n}{2 \pi} \Big(
 \ln \frac{16 \pi {\S_1}}{n}  - 1 +  \ln \sin \frac{\pi }{n} \Big)   +
  \mathcal{O}(\frac{1}{\mathcal{S}})   \ . \la{spik}
 \ee
 We observe that the subleading $\ln \sin \frac{\pi }{n} $
  term in  \rf{sj}   differs from the one in the spiky string  case
  one by an extra  factor of 2.
 This   difference  might  be attributed to  the fact
  that the subleading term is sensitive to the
  near-boundary (turning point of $\r$) region where the spiky string  has a cusp
   while
  the  present $S_1=S_2$ solution has a regular ``round arc'' shape.

The asymptotic   solution  of \rf{xoe}     in this  large spin  limit   is  given by
\begin{equation}
x=\cosh \rho \approx \sqrt{a_{-}} \cosh \big(\frac{\rK[\m]n}{\pi} \sigma\big)
\approx \  \cosh \rho_- \  \cosh (\kappa \sigma)
 \la{slo}
\end{equation}
At $\sigma=0$ the radial coordinate  $\r$  reaches its minimum  while at $\sigma= {\pi\ov n }$ it grows to infinity
  as $\kappa$ is large, i.e. the end-points of  $n$ arcs of the string approach the boundary of $AdS_5$.
Assuming the  initial condition $\theta(0)=0$ we obtain the approximate
solution for $\theta$
on the interval $ 0 < \s  < { \pi \ov n}$:
\begin{equation}
\tan \theta= \frac{\tanh \kappa \sigma}{\sinh \r_-} \ . \la{hh}
\end{equation}
Since $\kappa$ is large, the value of $\theta$
at the end of the arc is $\tan \theta=\frac{1}{ \sinh \r_- } =\frac{1}{\sqrt{a_{-}-1}}=\tan \frac{\pi m}{n}$,
which  matches the value  in (\ref{ddi}) in the limit  we are considering in this subsection
(see \rf{shu}).

Combining \rf{slo} and \rf{hh}  we can determine the  shape of the string   $\r(\te)$
\be
\cosh \r  = { \cosh \r_- \ov  \sqrt{  1 -  \sinh^2 \r_- \ \tan^2 \te }}  \label{slo1}
\  . \ee
 We conclude   that like   in the folded \ci{ft2}   and the spiky \ci{kt}    string cases
 the  solution in the large spin limit is represented by a combination  of
  $n$  arcs  stretching
 to the boundary  with each arc  described  by    a  simple analytic expression.

  In  section  3 we shall  rederive this
  asymptotic  solution  by starting with the assumption (valid in the large spin limit)
  that $\omega=\kappa$.


\renewcommand{\theequation}{3.\arabic{equation}}
 \setcounter{equation}{0}

\setcounter{equation}{0}

\section{Asymptotic solution:  relation  to  single-spin string }

Let us go  back to the system \rf{sqr}--\rf{con}  and try to  solve it  by assuming that
 $\omega_1=\omega_2=\kappa$. Then
\begin{equation}\la{ko}
\theta'=\frac{c}{\sinh^2 \rho}\ , \qquad \qquad \rho'^2= \kappa^2- \frac{c^2}{\sinh^2 \rho}
\end{equation}
This system  will not have a   finite-length  closed string solution.
We may, however,   relax the closed string condition,  find  its ``open-string'' solution
and then use it
to ``glue''  an asymptotic closed string  solution
 that in the $\k=\om \gg 1$   limit will  coincide
with the large spin  limit  of the
solution from the previous section.

\subsection{Solution for   $\omega_1=\omega_2=\kappa$}

The equation for $\rho(\s)$ in \rf{ko}
\begin{equation}\la{pp}
\rho'= \frac{\k }{\sinh \rho}\sqrt{\sinh^2 \rho- \sinh^2 \r_-}
\end{equation}
 has the following solution with  the initial condition  $\r(0)= \rr$
\begin{equation}
\cosh \rho (\s) =\  \cosh \r_-\  \cosh (\kappa \sigma) \ . \label{alk}
\end{equation}
Then
\begin{equation}
 \cot \theta(\s) =\ \sinh \r_-\ \coth (\kappa \sigma)  \ .  \label{lk}
\end{equation}
If $\s$ changes  in the interval  $0 < \s < \s_0 $ then  the maximal value of $\r$ is
given by $\cosh \rho_+ =  \cosh \r_-\  \cosh (\kappa \sigma_0)$.\foot{Note that  here $\te'$ never vanishes  so this  solution  cannot be interpreted as
a usual free open string.}
For example, in the previous section we had $ \sigma_0= { \pi \ov n}$
for a  half-arc of the string  and  $\k \gg 1$ in the large spin limit.
In this case $\rho_+ \to \infty$, i.e. the string stretches  all the way to the boundary
while $\te$ changes from  0 to  $ \arccot(\sinh \rr)$.

From \rf{ko} we get  the following equation for $\rho=\rho(\theta)$
\begin{equation}
\frac{d \rho}{d \theta}=\pm \frac{\sinh \rho}{\sinh \r_-}\sqrt{\sinh^2 \rho- \sinh^2 \r_-} \ , \ \ \ \ \ \ \
\sinh \r_-\equiv  \frac{c}{\kappa} \ . \la{iiu}
\end{equation}
Here   $\rho$ will  change  from the   minimal  value $\rho_-$ to  some maximal value.
The solution for $\theta(\r) $  on the interval $ \rr < \r < \infty$
with the initial condition $\te(\r_-) = 0$ is  (for the plus sign  choice in \rf{iiu})
\begin{equation}\la{qq}
\cot \theta =  \frac{\sinh \r_-\ \cosh \rho}{\sqrt{\cosh^2 \rho - \cosh^2 \r_-}} \ .
\end{equation}
Thus $0  < \te  < \te_0$,\  $\te_0 =\arccot( \sinh \rr )  < { \pi \ov 2}$.
In Figure 2 we   plotted
$\theta(\rho)$ for $\rr=1$.
\begin{figure}
\epsfig{file=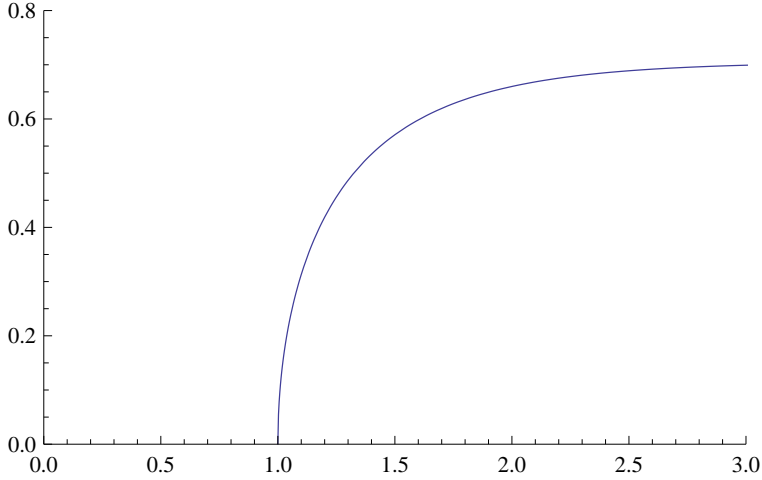, width=10cm}
\caption{Plot of $\theta(\rho)$ for $\rr=1$: from  $0$  at $\r=1$
and  to $\te_0 \approx 0.705$  at $\r=\infty$.}
\label{fig2}
\end{figure}

Joining together  two of such  stretches of string (with $\r$ changing from infinity to $\r_-$
and then  back to infinity while  $\te$ going from $-\te_0$ to 0 and to $\te_0$)\foot{Here $\bs=\k\s$
may be assumed to be changing from $-\infty$  to $+\infty$
(corresponding to $\s$ going from $-\pi$ to $\pi$ for a single fold  case
 and $\k \to \infty$).}
we get a single-arc open string solution (see Figure 3). Doubled version of
this arc might be interpreted\footnote{This interpretation makes sense if a two-spin bended  folded  string
solution can indeed  reach the  boundary, which is not a priori clear.} as
a  bended folded string  anticipated  in \ci{afrt} to be the large
spin limit of a   2-spin configuration  in $AdS_5$.

\begin{figure}
\epsfig{file=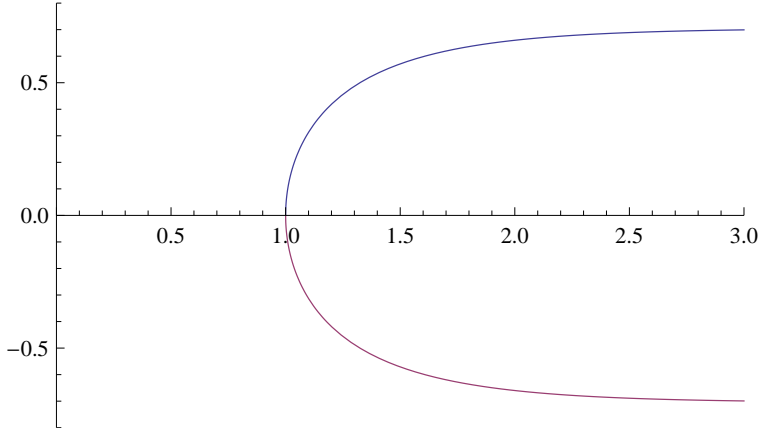, width=10cm}
\caption{String arc  stretching to infinity:
$\theta(\rho)$  changes  from  $\te(\infty)= -\te_0$
to $\te(\infty)=\te_0$  through $\te(\r_-)=0$. }
\label{fig22}
\end{figure}

\subsection{Relation to asymptotic limits of the
folded, spiky  and 2-spin  closed string solutions}

The  ``open-string'' or  ``half-arc''   solution  constructed above
is, in fact, equivalent to a  similar single-spin  solution
in $AdS_3$. This can be seen   by  writing it in terms of the $AdS_5$ embedding coordinates \rf{yq},\rf{uuu}.
The solution \rf{alk},\rf{lk} takes  the form
\be
&& Y_{05}= \cosh \r_- \ \cosh \bs  \  e^{i  \btau}, \qquad
\ \
Y_{12}= \sinh \rr \  \cosh \bs \  e^{i \btau }  , \qquad  \ \
Y_{34}= \sinh \bs \  e^{i \btau }, \quad \quad\label{dal}\\
&& \ \ \ \btau\equiv \k \tau \ , \ \ \
\ \ \ \bs \equiv \k \s   \la{jj}
\ee
Applying a  global  $SO(2,4)$ transformation with boost parameter
$\rr$    we  can transform \rf{dal} into an $AdS_3$ solution:
\begin{equation}
Y'_{05}= \cosh \bs  \  e^{i  \btau},
\quad \quad Y_{34}= \sinh \bs \  e^{i \btau }, \quad \quad
 Y'_{12}=0  \label{stra}
\end{equation}
For $\bs= \k \s$ changing   from $0$ to $\infty$ (for $\k \to  \infty$)  this
is the same  solution as found (for a single stretch of the string or  a  half-spike)
 in the large spin limit of the folded string  \ci{ft2}
 or  the spiky string  \ci{kt}.

 One  can give  a general proof that an open string solution in $AdS_5$
 described by the ansatz \rf{jk} with
 $\kappa=\omega_1=\omega_2$ can  be always
 $SO(2,4)$ transformed     into  an $AdS_3$  solution.
Starting with   the following ansatz for  the embedding coordinates\foot{One
may consider
also more  general
 solution with Y$_a=(Y_{05}, Y_{12}, Y_{34} )= z_a(\sigma+b \tau) e^{i \k \tau} $
  where $z_a$ are complex  and $\sigma$ is decompactified.
  One may  redefine $\sigma$ and $\tau$ preserving the conformal gauge
  to transform this into  Y$ _a=z_a(\sigma') e^{i \k' ( \tau' +  c \sigma')} $.
   Such a solution was  found  in an unpublished work
of  A. Irrgang and M. Kruczenski. It appears  to be  related to the one
 we discuss here  by  an $SO(2,4)$ transformation. }
\be \la{aqq}
Y_{01}=y_0(\sigma) e^{i \kappa \tau},  \qquad  Y_{12}= y_1(\sigma) e^{i \kappa \tau},
 \qquad Y_{34}= y_2(\sigma) e^{i
 \kappa \tau} \ee
we find (using the conformal gauge constraint)   that the equations of motion are
$y_a''- \kappa^2 y_a=0 \  (a=0,1,2)$, where the solutions should satisfy
$
y_0^2-y_1^2-y_2^2=1$.
 The most general solution   is $
y_a= A_a e^{\kappa \sigma} + B_a e^{-\kappa \sigma}$
and the constraints imply
$
A_0^2=A_1^2+ A_2^2,  \quad B_0^2=B_1^2+B_2^2,  \quad A_0 B_0-A_1 B_1- A_2 B_2=\frac{1}{2}
$.
These relations   determine  three parameters, while one additional
parameter can  be fixed   by a shift in $\sigma$, i.e.
 we are left with two free parameters. Since there are only two
 functionally-independent   terms $ e^{\kappa \sigma} $ and  $ e^{-\kappa \sigma} $
   this implies  that we can always set to zero,   say,
    $A_2$ and $B_2$  by an £$SO(1,2)$  transformation acting on $y_a$; this then   puts the
solution  into $AdS_3.$ \foot{For example,
we can first do rotation in the $(y_1,y_2)$ plane   to make $A_{2}=0$.
Then we get
$y'_0= A_0 e^{\kappa \sigma} + B_0 e^{-\kappa \sigma}$, \
$y'_1= A_0 e^{\kappa \sigma} + B'_1  e^{-\kappa \sigma}$, \
$y'_2=                                           B'_2 e^{-\kappa \sigma}$.
Then we can  boost in  $(0,1)$ plane
to make  $B'_1=0$,  getting
$y''_0= A_0 e^{\kappa \sigma} + B_0 e^{-\kappa \sigma}$, \
$y''_1= A_0 e^{\kappa \sigma}                                          $, \
$y''_2=                                           B_0 e^{-\kappa \sigma}$,
with $A_0 B_0 = \frac{1}{2}$.
Shifting  $\sigma$ by a constant we can set
$A_0 = B_0  ={ 1 \ov \sqrt 2}$,   i.e. end up with
$y_0 = y_1 + y_2 $, \
$y_1= { 1 \ov \sqrt 2} e^{\kappa \sigma}                                          $, \
$y_2=  { 1 \ov \sqrt 2}e^{-\kappa \sigma}$.
Finally,   we can  make an $SO(1,2)$ transformation that sets $\td y_2=0$:
$\td y_0= \cosh {\kappa \sigma}
 = \sqrt 2  [ y_0   - {1\ov  2}  (  y_1 + y_2)]    $, \
$\td y_1= \sinh {\kappa \sigma}
={ 1\ov \sqrt 2} ( y_1  -y_2)    $, \
$\td y_2= 0=  -  y_0    + y_1 + y_2      $.}


Since $Y_{05}$ and $Y_{12}$ in \rf{dal} are even under $\s \to - \s$
the global  boost  leading to  \rf{stra}  applies also to the full
arc  solution or its bended folded  string generalization  mentioned above.
This  transformation ``straightens up''  the folded string, i.e. removes  the bend
 making  string  pass through  $\r=0$  and thus making it   equivalent to large spin limit of the
 single-spin folded string.

Before the  $SO(2,4)$ rotation leading to \rf{stra}
 some of the ``non-Cartan''  components   of the $SO(2,4)$  generators are non-zero so the right
 semiclassical state interpretation
 of this  half-arc  solution  is that of a single-spin solution with spin equal to $S$.
 However, different ways of  gluing   single-arc solutions
 into a  closed string solution  which do not, in general, commute
 with $SO(2,4)$  rotations  may lead to inequivalent  solutions
 representing inequivalent  semiclassical string states
 once the strict large spin limit is relaxed.

 Indeed, the ``straight'' string  in \rf{stra} (for which $\te=0, \ \phi_1=0$ in \rf{yq})  can be
 glued with itself  into (an infinite spin limit of) the  folded string.
Applying a  different  global  $SO(2,4)$  boost
to \rf{stra}
one finds a different ``bended''
 form of the  single-spin  arc string; gluing together  $n$ such arcs
 gives \ci{kt} the asymptotic form of the $n$-spike solution of \ci{kru}.

\

Let  us  now discuss   more explicitly    how   to  reproduce the
asymptotic form of  the 2-spin solution \rf{slo},\rf{hh}
by starting with \rf{alk},\rf{lk}  or \rf{dal}.
We need to  glue together $2 n$ such segments (with $ 0 \leq  \s \leq \s_0={ \pi \ov n}$)
 to form an $n$-arc closed
 string  with end-points of the arcs approaching
  the boundary.
  Allowing  for the  winding number $m$   in  $\theta$,   the closed string condition is then
\begin{equation}
2 n\ \arccot (\sinh \r_-) = 2 \pi m\ . \la{gll}
\end{equation}
This implies a relation between $c$ and $\kappa$, or, equivalently,
fixes the minimal  value of $\r$
\begin{equation}
\cot \frac{\pi m}{n}= \sinh \rr =  {c \ov  \kappa}\  .
\label{aah}
\end{equation}
This  is the same   value (\ref{ahg}) found  by taking the large spin limit of the 2-spin solution
 in the previous section. Unless $c=0$ (when  $\rr=0$ we are   back to  the straight
 folded string case)
 we again  get the condition $2 m <n$, implying that the minimal values
 are $m=1,\ n=3$. The plot  of such solution  obtained  by gluing  3 arcs  of
 the type in  Figure 3
 is presented in Figure 4. Viewed as a limit of the solution plotted in Figure 1,
 the end-points of the 3  arcs  should be rounded. This is in contrast with
  the single-spin spiky string
 in $AdS_3$  \ci{kru}  where the spikes were present for any   value  of the spin.

\begin{figure}
\epsfig{file=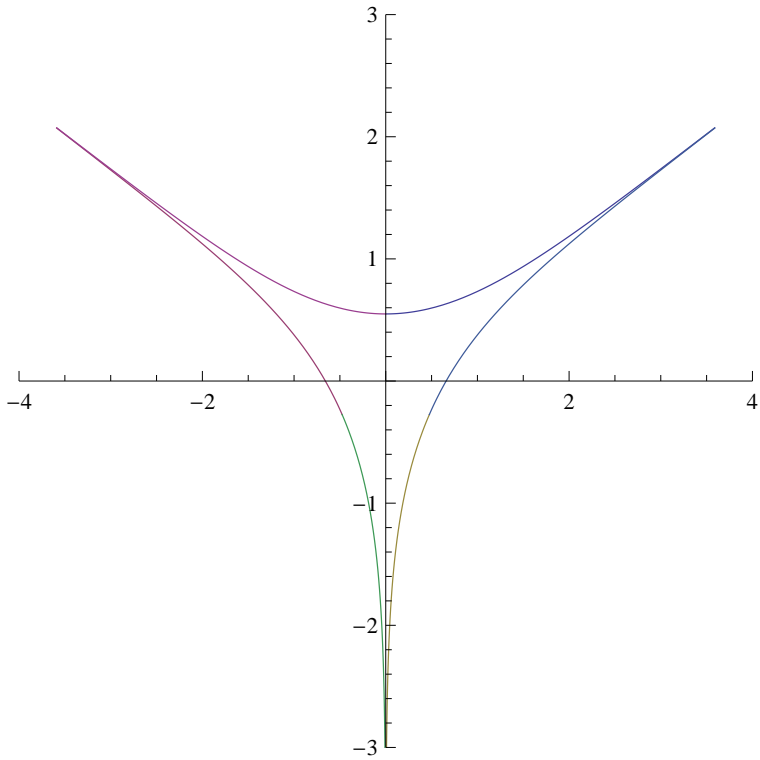, width=10cm}
\caption{Shape of $S_1=S_2$  string with  $m=1$, $n=3$  in the  $(\r,\te)$  plane.
The end-points are, in fact,  not cusps  but  rounded up as in Figure 1.
}
\label{fig3}
\end{figure}

Assuming that $\k \to \infty $,      we find  for  a half-arc solution  \rf{dal} \
 ($ 0 < \k \s  < { \pi \ov n} \k \to \infty $)
\begin{equation}
\mathcal{E}-\mathcal{S}= \frac{1}{4 \pi}\ln \mathcal{S}+...\ ,  \la{ot}
\end{equation}
where $\S$ is the total spin. This is
to be multiplied  by \ (i) 4 in the case of the  folded  single-spin string,
 (ii)  $2n$ in the case of the single-spin $n$-spike solution  and  also   (iii)
  $2n$   in the case of the  two equal spin
 $n$-arc solution of the previous section  (cf. \rf{sj}).

The folded  string is a special  case of the spiky string and represents the
lowest-energy state  for given  spin in the sector of one-spin solutions (or, in
particular,  in the $sl(2)$ sector of the  dual  gauge
theory). The  ``lightest'' non-trivial spiky string  with $n=3$
has    higher  energy  than the ground state
(folded string)  in the one-spin  set of states.
The lightest  non-trivial  two-spin solution
with   $m=1, \ n=3$   having
\begin{equation}
\mathcal{E}_{\rm min} -\mathcal{S}= \frac{3}{2 \pi}\ln \mathcal{S}+... \la{otd}
\end{equation}
should be representing the ``ground  state''
in the class  of   two-spin solutions with equal spins; the same
should  apply also to  the corresponding gauge
theory   states.\foot{We thank A. Rej for
 a discussion of this point.}

 It was  suggested  in \ci{kt} that for the single-spin spiky string
  the coefficient of $ n \ln S$ term  in the energy  should be
the same for  all $n$ to all  orders in the
${1 \ov \sql }$ expansion
 as it is determined by the same asymptotic  solution as in the folded string case.
 The above   conclusion about   universality of the  half-arc solution
 implies that the coefficient  of the  leading $ n \ln S$ term in  the
  $S_1=S_2$  string energy  should  also  be the same for any $n$ and any value of spin.
 Thus  the coefficient of   the  $n \ln \S$   term in the large spin expansion of
 the  string energy of  the 2-spin solution  of the previous section  (cf. \rf{sj})
 should be  given  by the  same universal  function of  the string
 tension, i.e.  the cusp anomaly function
 \begin{equation}
{E}-{S}= \frac{n}{2} f(\l) \ln {S}+ ... \  , \ \ \ \ \ \ \ \ \
f(\l)_{_{\l \gg 1}}  ={\sql \ov \pi}  + O( {1}) \ , \ \ \ \ \   n=3,4,....
\end{equation}


\subsection{Asymmetric gluing
}

Let us consider
possibility  of  gluing arcs
asymmetrically to  construct circular solutions  with unequal spins.
 We concentrate on
the minimal energy solution  with $m=1$  and number of arcs
 $n=3$.\footnote{Since arccot$ (\sinh \rho_{-}) \leq \frac{\pi}{2}$  to satisfy the gluing condition
 for $m=1$ we need at least $n=3$.}
 The asymptotic solution discussed below   should
 correspond to the asymmetric solution  of
 subsection 2.4 in the large $a_{+}$ limit.

We  would like  to glue
together   three different arcs  described by the asymptotic solutions (\ref{slo}),(\ref{hh})
\begin{equation}
\cosh \rho_i= \cosh \rho_{-i} \cosh \kappa \sigma, \ \ \ \  \ \cot \theta_i= \sinh \rho_{-i} \coth \kappa \sigma,
\ \ \ i=1,2,3
\end{equation}
with constant  parameters  $\rho_{-i}$  (minimal values of $\r_i$)
being, in general, different. To explicitly construct   the solution we need  to split  $0 < \sigma\leq 2\pi $
interval into  six  $\frac{\pi}{3}$ intervals.
The gluing condition (cf. \rf{gll})   is  $\Delta \theta = 2
 \sum_i  \arccot  (\sinh \rho_{-i}) = 2 \pi $, which   gives
\begin{equation}
\arctan (\sinh \rho_{-1})+\arctan (\sinh \rho_{-2})+\arctan (\sinh \rho_{-3}) = \frac{\pi}{2} \ .
\end{equation}
Thus only two out of three parameters  $\rho_{-i}$  are independent, e.g.,
\begin{equation}\la{rqq}
\sinh \rho_{-3}= \frac{1- \sinh \rho_{-1} \sinh \rho_{-2}}{\sinh \rho_{-1} + \sinh \rho_{-2}}
\end{equation}
An example  of resulting solution is  shown in Figure  5.
\begin{figure}
\epsfig{file=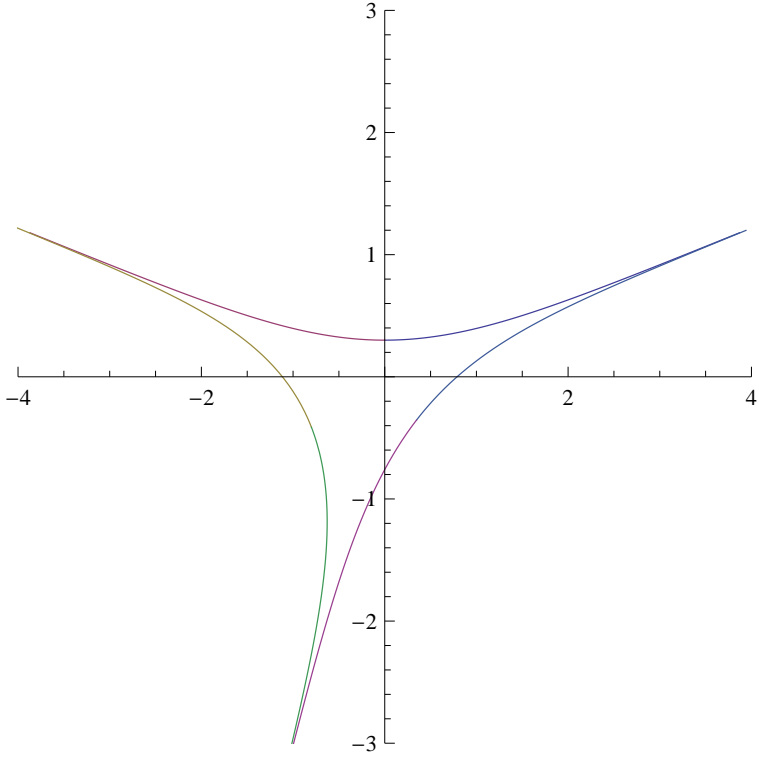, width=10cm}
\caption{Asymmetric circular string with  $m=1$, $n=3$, $\rho_{1-}=0.3$,
 $\rho_{2-}=0.9$, $\rho_{3-}=0.5$, $\frac{S_1}{S_2}=0.64$ in the
  $(\r,\te)$  plane.}
\label{fig33}
\end{figure}

Computing  the spins  at  leading order in  large $\k$ expansion
we get
\begin{equation}
\mathcal{S}_1 \approx
 \frac{e^{\frac{2\pi}{3}\kappa
 }}{16 \pi}  h_1(\rho_{-1},\rho_{-2}),\ \ \ \ \ \ \ \ \ \mathcal{S}_2
\approx  \frac{e^{\frac{2\pi}{3}\kappa
}}
{16 \pi}h_2(\rho_{-1},\rho_{-2})
\end{equation}
where  we used \rf{rqq} and
the functions $h_1, h_2$ are,  in general, different\foot{They are equal only in the
symmetric case,
 which corresponds to
$\sinh \rho_{-1}=\sinh \rho_{-2}=\sinh \rho_{-3}=\frac{1}{\sqrt{3}}$.}
 (their explicit form is
somewhat complicated).
Hence in the large $\k$  case $\S_1, \S_2 \gg 1$ while
$\S_1 \ov \S_2$ is fixed.
The energy is then
\begin{equation}
\mathcal{E}-\mathcal{S}_1-\mathcal{S}_2= \k \approx \frac{3}{2 \pi}\ln \mathcal{S}_1  + O(1)
=  \frac{3}{2 \pi}\ln ( \mathcal{S}_1 + \mathcal{S}_2)      +  O(1)  \ .
\end{equation}
The coefficient of the  leading $\ln \S$ term is thus independent of $ \S_1 \ov \S_2$, i.e.  is
the same as in the symmetric gluing case.

 A puzzling feature of the resulting solution is that
 some of the ``off-diagonal'' components  of the $SO(2,4)$
  angular momentum are apparently non-zero, suggesting  that the gluing procedure should be more subtle
  (since otherwise this solution does not represent a highest-weight state and thus one
  should be able to rotate it
  into a simpler solution).
 One option is  that this glued solution is not actually a
 limit of any  regular   closed string solution. Indeed,
 as follows from \rf{iiu}, the derivative $ d \rho \ov d \theta$  for large $\rho$
 goes as $ \sinh^2 \r \ov \sinh \r_-$; since it depends on $\r_-$
 the derivatives  from the ``left'' and from the ``right''   at the  end of the string
 will not match  and thus  there will be a cusp.


\renewcommand{\theequation}{4.\arabic{equation}}
 \setcounter{equation}{0}

\setcounter{equation}{0}

\section{``Circular'' solution with $(S_1,S_2)$: \  $\kappa=\omega_2 \not=\om_1$}

Let us now  present  an explicit 2-spin   solution with unequal spins  by considering the second
special case in \rf{eee}, i.e.
 $\kappa=\omega_2 \neq \omega_1$.  In this case the string will   again have
  a topology of a
  circle. The large-spin limit in which the string  can reach  the boundary will be possible only for
  $S_1=S_2$. In this case  the  asymptotic solution can be  again glued out of parts
   equivalent to the one  discussed in section 3.

 \subsection{Constructing solution}

To integrate the  system of equations for $\kappa=\omega_2$
    it is useful to
 change the  coordinates in \rf{io},\rf{uuu}  from $(\r,\te)$ to
$(\z,\psi)$ defined according to
\be
\cosh \rho= \cosh \z \  \cosh \psi\ , \ \ \ \ \ \ \ \ \
\sin \theta= \frac{\cosh \z\ \sinh \psi}{\sqrt{\cosh^2 \z \ \sinh^2 \psi  +    \sinh^2 \z}}
\ .   \label{amk}
\ee
Then the  embedding coordinates in \rf{yq},\rf{uuu}  take the following  form
\be
y_0=  \cosh \z \  \cosh \psi \ , \ \ \ \
y_1 = \sinh \z \ , \  \ \
y_2 =  \cosh \z \  \sinh \psi \  \la{u}
\ee
and the  \ads metric \rf{io}   becomes
\begin{equation}
ds^2=- \cosh^2 \z \ \cosh^2 \psi\ dt^2 +   {d\z^2}+  \cosh^2 \z \ d \psi^2
 +   \sinh^2  \z\ d \phi_1^2 + \cosh^2 \z \  \sinh^2 \psi\  d \phi_2^2
\end{equation}
Assuming the ansatz  ($\om_1\equiv \om$)
\begin{equation}
t=\phi_2=\kappa \tau, \quad \phi_1=\omega \tau, \quad
  \quad \z=\z(\sigma), \quad \psi=\psi(\sigma) \ ,
\end{equation}
the  equation  for $\psi$ can be easily integrated (cf. \rf{sqr})
\begin{equation}
 \psi'= { c \ov  \cosh^2 \z }  \ .  \la{pip}
\end{equation}
When $c=0$  we have $\psi= \psi_0$=const and the resulting solution is
related by a global $SO(2,4)$ boost to  the single spin  folded string solution (cf. \rf{yq},\rf{u}).
For $c\not=0$ we  have  $\psi'\not=0$  and so the string's   shape cannot be
of bended folded type, i.e. it should  be of circular type (see \ci{afrt}  and section 1).

The  conformal constraint gives the following  equation for $\z$
\be
x'^2 = \k^2  (1+ x^2)^2 - c^2  - \om^2 x^2 (1+ x^2)  \label{zzz}
\ , \ \ \ \ \ \ \ \ \
x\equiv y_1= \sinh \z  \ ,
\label{egg}
\ee
which is very similar to the one we had in  \rf{yyy}.
The  equations (\ref{yyy}) and (\ref{zzz})  are related by the analytic continuation
$x \rightarrow  i x$,\  $\om \rightarrow \kappa$.
Indeed,  the equation for $x\equiv y_0=\cosh \rho $  in \rf{yyy}  we had in the $\om_1=\om_2$ case in
the present $\k = \om_2$ case is  replaced by
the equation for $x\equiv y_1=\sinh \z  $.\foot{This effective replacement is implied by the general analysis
of equations for $y_a(\s)$ in \ci{afrt}.}
In this sense  $\z$ is now playing  the role of $\r$ and
 the subsequent analysis of the  solution  of  \rf{egg} will be  similar to the one in section 2.
Eq. \rf{egg} can be written  in the same way as \rf{xoe}:\foot{In this section we shall use
similar notation  for the  parameters  as in section 2 but their values
 will be  of course  different as
the  two solutions are  different.
}
\begin{equation}
x'^2=(\om^2-\kappa^2)(x^2-a_{-}) (a_{+}-x^2) \ , \la{koy}
\end{equation}
where
\begin{equation}\la{koyy}
a_{\pm}=\frac{2 \kappa^2-\om^2 \pm \sqrt{\om^4-4 c^2(\om^2-\kappa^2)}}{2 (\om^2-\kappa^2)}
\end{equation}
If  $a_{\pm}>0$  (as we shall assume in this section)
then  $\sqrt{2} \kappa > \om > \kappa$  and  $c^2 >\kappa^2$.
\footnote{For $\om<\kappa$ there are no closed string solutions.
If $c=\kappa$ we get $a_{-}=0$ and $a_{+}
=\frac{2 \kappa^2-\om^2}{\om^2-\kappa^2}.$ In this case the string can reach the center of
$AdS_5$ and one may expect to get typical near flat space expressions. However,
for $c=\kappa$ the motion is not periodic in $\s$:  we get  $x=\sinh \chi \sim
 \tanh \k \sigma$, i.e. there is only the ``open string'' solution
resembling   the giant magnon \ci{hm} one.}
 Then $\chi$ changes from a  minimal to a maximal value.
The case when $a_+ >0 $ and $a_- < 0 $  will be discussed in Appendix.
Some useful relations are
\begin{equation}
\frac{\om^2}{\kappa^2}=\frac{a_{+}+a_{-}+2}{a_{+}+a_{-}+1}\ , \quad \quad \om^2-\kappa^2=\frac{\kappa^2}{a_{+}+a_{-}+1}\ ,
 \quad \quad \frac{c^2}{\kappa^2}=1+\frac{a_{+} a_{-}}{a_{+}+a_{-}+1} \label{lbv}
\end{equation}
It is convenient to introduce the following parametrization of $a_\pm$   ($\m > \n$)
\begin{equation}\la{lbvv}
\m=\frac{a_{+}-a_{-}}{a_{+}}, \quad \quad \n=\frac{a_{+}-a_{-}}{a_{+}+1}, \quad \quad a_{+}=\frac{\n}{\m-\n},
 \quad \quad a_{-}=\frac{\n(1-\m)}{\m-\n}
\end{equation}
Solving \rf{koy}  with the initial condition $x(0)=\sqrt{a_{-}}$ on
an interval $0 < \s  \leq  \s_0=\frac{\pi}{n}$
we find
\begin{equation}
x=\sinh \chi=\pm \frac{\sqrt{a_{-}}}{\dn[\frac{\omega \sqrt{a_{+}}}{\sqrt{a_{+}+a_{-}
+2 }}\sigma,\m]}
 \label{sla}
\end{equation}
where  $x(\sigma= \frac{\pi}{n}   )=\sqrt{a_{+}}$.
We will then need  to  glue together $2n$ of such segments to form a closed string.
 Because the period of the  $\dn(z,\m)$
 Jacobi function is $2 \rK[\m]$ we need to satisfy
\begin{equation}
\om \sqrt{\frac{a_{+}}{a_{+}+a_{-} +2 }} = \frac{\rK[\m] n}{\pi}
\end{equation}
This equation along with (\ref{lbv}) can be used to eliminate one constant out of $\kappa, \om, c$.
The corresponding equation for $x=x(\psi)$ that follows from \rf{pip} and \rf{koy} is
\begin{equation}\la{pssi}
\bigg(\frac{dx}{d \psi} \bigg)^2=
\frac{1}{(a_{+}+1)(a_{-}+1)}(1+x^2)^2 (x^2-a_{-})(a_{+}-x^2)
\end{equation}
Taking the square root there are two branches. Using the negative branch
 and the initial condition that at $\s=0$ where $x= \sqrt{a_{-}}$ we should have
 $\psi=0$ (i.e.   $\psi(x=\sqrt{a_{-}})=0$)
we obtain the following solution
\begin{equation}
\psi(x) =\frac{\sqrt{a_{-}+1}}{\sqrt{a_{+}(a_{+}+1)}}\bigg(\Pi[\n,\m]-\Pi[\n,\arcsin
\sqrt{\frac{a_{+}-x^2}{a_{+}-a_{-}}},\m]\bigg)  \label{pss}
\end{equation}
At $\sigma=\frac{\pi}{n}$  where  $x=\sqrt{a_{+}}$ we have
\begin{equation}
\psi(\sqrt{a_{+}})\equiv \psi_0=\frac{\sqrt{a_{-}+1}}{\sqrt{a_{+}(a_{+}+1)}} \Pi[\n,\m] \la{por}
\end{equation}
Thus $\psi$ increases from $0$ to $\psi_0$.
The solution for the positive branch   has the opposite sign, decreasing from 0 to $-\psi_0$.
To get $\psi=\psi(\sigma)$ one may  plug the solution for $x(\s)$ (\ref{sla}) into  \rf{pss}.


\subsection{Closed-string  condition, energy and spins}

To construct a closed string solution  that will have a circle shape
we need to  glue
together  $2n$ of the  above string segments.
We should  first  express  the string segments
 obtained above in  $(\chi,\psi)$  coordinates
in  terms of $(\rho, \theta)$ coordinates (with $0 < \theta \leq 2\pi $), and
 then glue them  together
 in a similar way as was done in section 2.

When  $\s$ changes from $0$ to $\s_0= { \pi \ov n}$, i.e.
when     $x=\sinh \chi$ changes from $\sqrt{a_-}$ to $\sqrt{a_+}$
and $\psi$ changes from $0$ to $\psi_0$ in \rf{por}
we get, according to \rf{amk}, that   $\te$ changes from 0 to $\te_0$ with
\begin{equation}
\sin \te_0= \frac{ \sinh \psi_0}{\sqrt{ \sinh^2 \psi_0+ {a_{+}\ov 1+a_{+}}}}
\label{cond4}
\end{equation}
 To get  a closed  string with a  circular shape in $\te$ we need to demand
($m$ is a  winding number)
\be \te_0 = { \pi m \ov n }  \  , \ \ \ \ \ \ \ \ \ \ 2m < n    \ .  \la{ttt}\ee
We found   (using numerical  methods)
that gluing  $2n$ segments  into  such closed-string solution
  is indeed possible, and also checked that all non-Cartan $SO(2,4)$    charges
   then vanish,   while in general  $S_1 \neq S_2$.
For  $m=1$, $n=3$ the  resulting  string shape  is shown in Figure 6 in polar
coordinates $(\rho, \theta)$.
As $a_{+}$ gets larger, the spins and $\rho$ grow.
We found that to have a  limit when the string touches
 the boundary ($\kappa \approx \omega$) is possible only when  the two spins are
 actually  equal
  $S_1=S_2$  (see below).

\begin{figure}
\epsfig{file=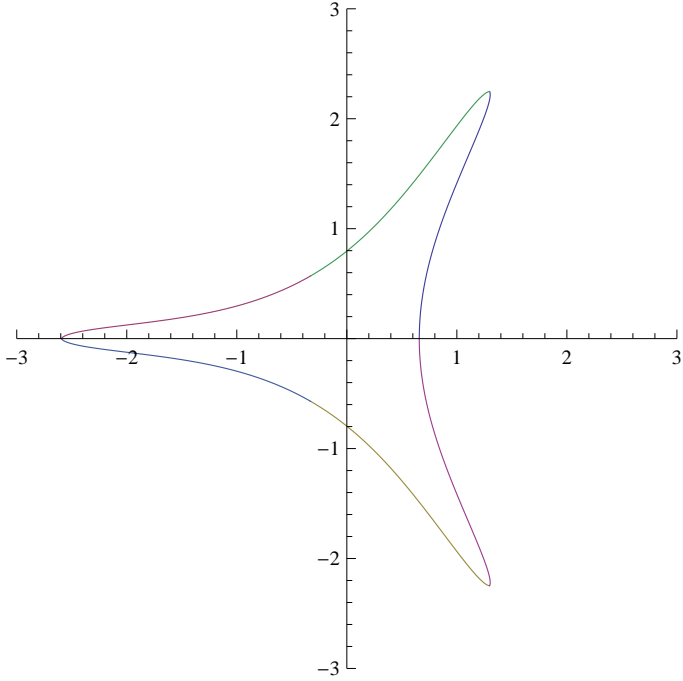, width=9cm}
\caption{Shape of $(S_1,S_2)$ string for $m=1$, $n=3$ in polar coordinates $(\rho, \theta)$
for $a_{-}=0.5$, $a_{+}=11.095$. In  this case $\frac{S_1}{S_2}=1.039$.}
\label{fig4}
\end{figure}

Let us describe  the gluing procedure in more detail.
Given $\chi(\s), \psi(\s)$  in  \rf{sla},\rf{pss}
we may use \rf{amk} to compute the corresponding   $\bar \r(\s), \bar \te(\s)$
for $0 < \s \leq {\pi \ov n}$.
To find the full closed string solution  $\r(\s), \te(\s)$ defined
on  $0 < \s \leq {2\pi}$   we attach together
 $\rho(\sigma)$ intervals  as in section 2  while $\theta$ is obtained by combining
its  values on separate  intervals as follows
\begin{equation}
\theta  =\left\{
             \begin{array}{ll}
               (k-1) \theta_0+\bar \theta(\sigma), & \frac{(k-1)\pi}{n}\leq \sigma \leq \frac{k \pi}{n}, \ \ \ k=1,3,5, ..., 2n-1 \\
               k \theta_0-\bar \theta(\sigma), & \frac{(k-1)\pi}{n}\leq \sigma \leq \frac{k \pi}{n}, \ \ \ k=2,4,6,..., 2n
             \end{array}
           \right.
\end{equation}
 $\theta(\sigma)$ defined in this way has period $\frac{2 \pi}{n}$.
 The energy and the spins  are found from  \rf{enk} with $\om_1=\om, \ \om_2=\k$. They thus
 satisfy $ \frac{\mathcal{E} - \mathcal{S}_2  }{\kappa}-\frac{\mathcal{S}_1}{\omega}=1$.
Since the energy  is given by an  integral of a function of $\r$ only,
 it  may  be computed directly using the expressions for  $\chi(\s), \psi(\s)$ in
\begin{equation}
\mathcal{E}=\frac{2 n \kappa}{2\pi}\int_0^{\frac{\pi}{n}} d \sigma \cosh^2 \chi \cosh^2 \psi\ . \la{jp}
\end{equation}
The two spins are given by

$
\mathcal{S}_1= \frac{\om}{2 \pi}\sum_{k=2,4,...}^{2 n}\big(\int_0^{\frac{\pi}{n}}d \sigma \sinh^2
 \rho(\sigma) \cos^2 [(k-2)\theta_0+\bar\theta(\sigma)]+ \int_{\frac{\pi}{n}}^{\frac{2\pi}{n}}d \sigma
  \sinh^2 \rho(\sigma) \cos^2 [k\theta_0-\bar \theta(\sigma)]\big)
$

$\mathcal{S}_2= \frac{\kappa}{2 \pi}\sum_{k=2,4,...}^{2 n}
\big(\int_0^{\frac{\pi}{n}}d \sigma \sinh^2 \rho(\sigma) \sin^2 [(k-2)\theta_0+\bar\theta(\sigma)]
+ \int_{\frac{\pi}{n}}^{\frac{2\pi}{n}}d \sigma \sinh^2 \rho(\sigma) \sin^2
[k\theta_0-\bar\theta(\sigma)]\big)  $

Like \rf{jp} these integrals  are  complicated (it does not seem possible
to express  them   in terms of elliptic functions).
 The only integrals that  simplify are those along the first
  segment since in that case we  can   map  back to the $\chi,\psi$ functions
\begin{equation}
\int_0^{\frac{\pi}{n}}d \sigma \sinh^2 \rho \cos^2 \theta = \int_0^{\frac{\pi }{n}}d \sigma \sinh^2
 \chi= \frac{\pi a_{+}}{n} \frac{\rE[\mu]}{\rK[\mu]}
\end{equation}
\begin{equation}
\int_0^{\frac{\pi}{n}}d \sigma (\cosh^2 \rho - \sinh^2 \rho\ \sin^2 \theta )=
 \int_0^{\frac{\pi}{n}}d \sigma \cosh^2 \chi= \frac{\pi}{n}\big(a_{+} \frac{\rE[\mu]}{\rK[\mu]}+1\big)  .
\end{equation}
The above expressions  produce, in principle, the functions
$\E=\E(a_+, a_-, n),$ \  $\S_1=\S_1 (a_+, a_-, n),$ \ $\S_2=\S_2 (a_+, a_-, n)$.
The condition \rf{ttt} with $\te_0$ given by \rf{cond4},\rf{por}
gives one relation between $a_+,a_-,n$ and integer $m$. This implies a relation between
$\S_1, \S_2,n$ and $m$. As a  result, one may
determine (at least numerically)    the energy as a function of the two spins:   $\E=\E(\S_1,\S_2, n)$.

From a numerical analysis   we concluded
that for an  arbitrary  value of $\S_1 \ov \S_2 $, the solution for $a_{+}$ is bounded, and the spins cannot be large.
It is only if  $\S_1=\S_2$ that we can have a solution with
 large $a_{+}$ (i.e.  large $\chi$ and thus large $\r$),
 so that the string  may  touch  the boundary and the spins can  take  large  values.

\subsection{Large spin limit}

Let us consider the case   when  the limit of $a_{+} \gg 1 $
is possible.
 This limit  corresponds to  $\n\approx \m \approx 1$  when
\begin{equation}
\omega \approx \kappa \approx \frac{n}{2\pi}\ln \frac{16}{1-\m} \gg 1   \ .
\end{equation}
The solution for $x=\sinh \z$ in \rf{sla} becomes
\begin{equation}\la{kj}
\sinh \z= \sqrt{a_{-}} \cosh ( \frac{\rK[\m] n}{\pi}\sigma) = \sqrt{a_{-}} \cosh (\kappa \sigma)
\end{equation}
Here we again relax the periodicity condition in $\s$ and consider only one interval
$ 0  < \k \s < \k { \pi \ov n }  \gg 1 $.
Then the equation \rf{pssi} for $\psi$  leads to
\begin{equation}
\tanh \psi= \frac{1}{\sqrt{1+a_{-}}}\tanh (\kappa \sigma)
\end{equation}
In this limit we can use the following approximation for $\Pi[\n,\m]$ for $\n<\m$ \foot{This
 is different
from \rf{ooo} as the parameters   here are different.}
\begin{equation}\la{rtw}
\Pi[\n,\m]=\sqrt{\frac{\n}{(1-\n)(\m-\n)}} {\rm arcsinh} \sqrt{\frac{\m-\n}{1-\m}} \ ,
\end{equation}
implying that
\begin{equation}
\sinh \psi_0= \sqrt{\frac{\m-\n}{1-\m}}  \ .
\end{equation}
The condition (\ref{cond4})  becomes
\begin{equation}
\cos \frac{\pi m}{n}= \sqrt{\frac{1-\m}{1-\n}}  \label{qqp}
\end{equation}
Here   $2 m<n$, i.e.  the minimal choice is  again
 $m=1$,  $n=3$.
Equation (\ref{qqp}) implies
$a_{-}={\cot^2 \frac{\pi m}{n}}$.
 Using  (\ref{amk}) we  then conclude  that the resulting asymptotic solution
 is the same as found in section 2.3 (cf. \rf{slo},\rf{hh})
\be
&& \cosh \rho = \cosh \r_- \ \cosh (\kappa \sigma ) \ , \ \ \ \ \ \ \ \ \ \
\cosh \r_- = \frac{1}{\sin \frac{\pi m}{n}}   \la{roq} \\
&& \tan \theta =   \tan \te_0  \  \tanh ( \k \s)    \ ,  \ \ \ \ \ \ \ \
 \tan \theta_0=  \tan \frac{\pi m}{n} = { 1 \ov \sinh \r_-}  \ .
\la{wwq}
\ee
 As we have seen in section 3,
starting with such asymptotic   solution  where   string is  stretching  towards
 the boundary we may glue such  arcs  together  to get a closed string solution
with two equal spins $\S_1=\S_2\gg 1 $ and $\ln \S$ scaling of the energy.
If we would try  to continue   this asymptotic solution to finite values of the spins
we would end up with the solution of section 2.1.

\renewcommand{\theequation}{5.\arabic{equation}}
 \setcounter{equation}{0}
\section{Comments on  general solutions with $\S_1 \not=\S_2\gg 1 $}

Let  us now   relax the  assumptions  about  the frequencies like in \rf{eee}
and try to determine  the  general properties of a solution that may
allow the large-spin limit, i.e.  for which parts  of the  closed string may stretch towards
the boundary ($\r \to \infty$). This may be  either a bended  folded string \ci{afrt}
or a circular string of the type
described in the previous sections.

\subsection{Universal $\ln S$ scaling}

Let us start with
 the string equations of motion and the conformal constraint (their first integral)
for  the ansatz  \rf{jk}   with  generic $\k, \omega_1,  \omega_2$: \foot{More general ansatz
for infinite (open) strings ending on  the boundary was considered in \ci{kruu}.}
\be
&& (\theta' \sinh^2 \rho)'=(\omega_1^2-\omega_2^2) \sin \theta \cos \theta\ \sinh^2 \rho \la{ll} \\
&& \rho''- \cosh \rho\ \sinh \rho\ (\kappa^2 + \theta'^2- \omega_1^2 \cos^2 \theta -\omega_2^2 \sin^2 \theta)=0
\label{sys}\\
&&\rho'^2 - \kappa^2 \cosh^2 \rho +\sinh^2 \rho\ \theta'^2 + \omega_1^2 \sinh^2 \rho\ \cos^2 \theta  + \omega_2^2 \sinh^2
 \rho\ \sin^2 \theta
=0
\la{hl}\ee
Being interested in  solutions  which  have parts stretching towards large   $\r$, let us
focus on such  asymptotic region  where
we may set  $\sinh \rho \approx \ha {e^{\rho}}$. Then  eqs. \rf{ll},\rf{hl}  reduce to
\be
&&(\theta' e^{2 \rho})'= (\omega_1^2-\omega_2^2) \sin \theta\ \cos \theta\ e^{2
 \rho}  \label{hst}\\
 &&
\rho'^2 - \kappa^2(1+ \frac{e^{2 \rho}}{4}) +
\frac{e^{2 \rho}}{4} \theta'^2 +\frac{ e^{2 \rho}}{4}(\omega_1^2 \cos^2
\theta + \omega_2^2 \sin^2 \theta)=0  \la{ts}
\ee
Let us  rescale $\s$  by $\kappa$, i.e.  introduce $\bar \s \equiv  \k \s$  and also define
 $w_i\equiv \frac{\omega_i}{\kappa}$. Then (now prime will be  derivative over $\bar \s$)
\be
&&(e^{2 \rho} \theta')' = (w_1^2-w_2^2) e^{2 \rho} \sin \theta\ \cos \theta \\
&&
4(\rho'^2 -1)- {e^{2 \rho}}+ {e^{2 \rho}}\theta'^2 + {e^{2 \rho}}(w_1^2 \cos^2
\theta + w_2^2 \sin^2 \theta)=0 \ . \ee
To get $\ln S$  behaviour of the energy at large spin we shall assume that $\k \gg 1$ so
that $\bar \s$ can take
large values  and
will  require  $e^{\rho}$ to increase exponentially  with  $\bar \s$.
Setting  $e^\r =u(\bar \s)e^{\bar \s }$  we get
\be
&&2 \theta' + 2 u \theta' + u^2 \theta''= (w_1^2-w_2^2) u \sin \theta \cos \theta
\\
&&
4\frac{u'^2+2 u u'}{u^2}-
{u^2 e^{2 \sigma}}+{u^2 e^{2 \sigma}}\theta'^2+{u^2 e^{2 \sigma}}(w_1^2 \cos^2 \theta+
 w_2^2 \sin^2 \theta)=0
\ee
These equations  are not readily solved but we  observe that
$\mathcal{E}- \S \approx  \frac{\k}{2 \pi}  \gg 1 , $\
where $\S =\mathcal{S}_1+\mathcal{S}_2$ and
 $\k$  plays the role of  a  cutoff on   $\bar \s$.
As in the previous sections,   the asymptotic solution reaching
the boundary may be built out of several
segments. For  such a segment
 for any   $u(\bar \sigma)$  at the leading order we have
$\k\approx  \frac{1}{2}\ln \mathcal{S}$  and thus
\begin{equation}
\mathcal{E}-\mathcal{S}= \frac{1}{4 \pi}\ln \mathcal{S}+...  \label{oan}
\end{equation}
To get a closed string solution we would  need to combine
together several such segments;  we need at least  four of them
 as in the case of the folded string.

The general conclusion appears to be   that for spinning strings with two  large spins
and  having minimal energy for given spins so that only
  parts of them are reaching the boundary
the $\ln \S$ behaviour of the energy in the large spin limit appears
to be {\it universal}.
 It remains an open problem to systematically
classify  such  string   solutions
with {unequal} ($S_1 \not= S_2$)  spins   which may be  taken to be large   with
their ratio  fixed.

\bigskip

\subsection{$S_1\not= S_2$  deformation of solution of Section 2 }

Let us now discuss how one may construct an example of a solution  with
$E \sim \ln S$ and $S_1 \not=S_2$  by perturbing the   equal-spin circular solution of section 2.
Let us go back to eqs. \rf{ll} and \rf{hl} writing them in terms of
the  rescaled parameters
$w_i=\frac{\omega_i}{\kappa}$ and  $\bar{\sigma}=\kappa \sigma$
\be
&& (\theta' \sinh^2 \rho)'=(w_1^2-w_2^2) \sin \theta \cos \theta\ \sinh^2 \rho \label{asa}
\\
&&\rho'^2 - \cosh^2 \rho +\sinh^2 \rho\ \theta'^2 +  \sinh^2 \rho\ (w_1^2 \cos^2 \theta  + w_2^2 \sin^2 \theta)
=0
\label{sys1} \ee
and  now  taking  $w_1=w$ and $w_2=w-\epsilon$ with  $\epsilon \ll 1 $.
Thus we will have $1 \leq w_2^2 \leq w_1^2$.
At leading order in small  $\epsilon$ expansion when
$w_1=w_2$  when
this  system of equations decouples  it was already   solved  in Section 2.
Expanding near this solution $\rho_0, \theta_0$  given in  (\ref{sol1}),(\ref{sol2})
with  (see (\ref{opii}))
\begin{equation}
w^2= \frac{a_{+}+a_{-}-1}{a_{+}+a_{-}-2}
\end{equation}
we have
\begin{equation}
\rho(\bar{\sigma},a_{\pm})=\rho_0(\bar{\sigma},a_{\pm}) + \epsilon \rho_1(\bar{\sigma},a_{\pm}),
 \ \ \ \ \ \ \ \theta(\bar{\sigma},a_{\pm})=\theta_0(\bar{\sigma},a_{\pm})
+ \epsilon \theta_1(\bar{\sigma},a_{\pm})  \label{alp}
\end{equation}
where from \rf{asa},\rf{sys1}  the linear perturbations should  satisfy
\be
&& \sinh \rho_0\big[2 (\rho_1' \theta_0' + \rho_0' \theta_1')\cosh \rho_0 + \theta_1''
 \sinh \rho_0\big] - 2 w  \sin \theta_0 \cos \theta_0 \ \sinh^2 \rho_0 =0 \label{sy1}
\\
&& \rho_0' \rho_1'+\rho_1 \sinh \rho_0 \cosh \rho_0 (w^2-1+\theta_0'^2)+\sinh^2
  \rho_0 (\theta_0' \theta_1'-w \sin^2 \theta_0) =0
\label{sys2} \ee
We can then solve the second equation for $\theta_1'$ in terms of $\rho_1, \rho_1'$ and plug
it into the first. This gives  the following equation for $\rho_1$
\begin{eqnarray}
&&  A \r''_1 + B \r'_1 + C\r_1 + D  =0 \la{khh}  \\
&& A=- 2 \rho_0''\theta_0'\ , \ \ \ \ \
B= \theta_0'  \big[(1-w^2+\theta_0'^2)\sinh (2 \rho_0)-2 \rho_0''\big]
+ 2\theta_0''  \rho_0' \no \\
&&C=  \theta_0'' (w^2-1+\theta_0'^2) \sinh (2 \rho_0)       \ , \ \ \ \ \ \
D=   2w\theta_0'  \rho_0' \sin^2 \theta_0 \sinh (2 \rho_0)
 -2 w  \theta_0''  \sin^2 \theta_0 \sinh^2 \rho_0
 \no
\end{eqnarray}
Given  the complicated form of the functions
$\rho_0, \theta_0$ we cannot solve   for $\rho_1$
but we have  checked   numerically  that eq. \rf{khh}
does have a solution with trivial initial conditions. \foot{
This equation can  be simplified by
by considering the  large spin limit when the string touches the boundary ($a_{+} \rightarrow \infty$).
 Then  $\k  \to \infty $, i.e.
  the range of $\bar{\sigma}$ is $0 \leq \bar{\sigma} < \infty$.
 We
 found in Section 2 that in this limit  (we  use
 $00$  subscript to denote the asymptotic form of the leading-order  $S_1=S_2$ solution)
$w=1,  \ \ \cosh \rho_{00} = \cosh \rho_{-} \cosh \bar{\sigma}, \ \ \cot \theta_{00}
=\sinh \rho_{-} \coth \bar{\sigma} $.
Then
\begin{eqnarray}
&& \frac{1-a_{-} \cosh^2 \bar{\sigma}}{\sinh \bar{\sigma}}\ \rho_{10}''-2 a_{-}  \cosh \bar{\sigma}\ \rho_{10}'
 - \frac{2 (a_{-}-1)}{[a_{-} \cosh (2 \bar{\sigma}) +a_{-}-2]\sinh \bar{\sigma} }\ \rho_{10} \nonumber\\
&& +\  2 \sqrt{a_{-}} \sqrt{a_{-}\cosh^2 \bar{\sigma}-1} \sinh( 2 \bar{\sigma}) =0 \no
\end{eqnarray}
This equation  still seems hard to solve analytically. Numerically we  again find a  solution
 with the initial conditions $\rho_{10}(0)=0$, $\rho'_{10}(0)=0$.}
It is plotted   in Figure 7.
\begin{figure}
\epsfig{file=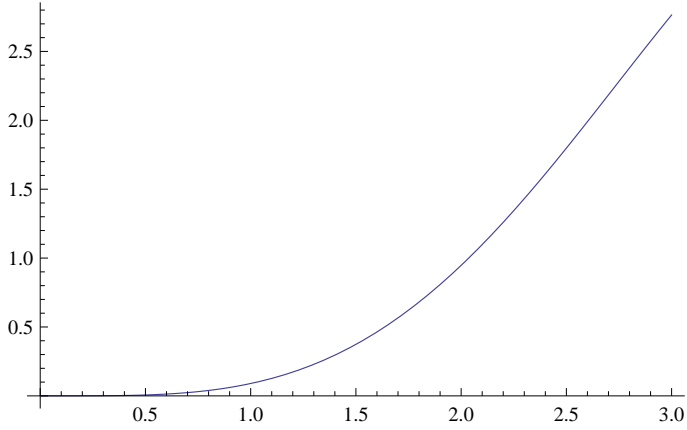, width=9cm}
\caption{ Solution for $\rho_{1}(\bar{\sigma})$ for $a_{-}=1.5$, $a_{+}=6.98$ with
 satisfying  $\rho_1(0)=0$, $\rho'_1(0)=0$.}
\label{fig44}
\end{figure}
Using the solution for $\rho_{1}$ one can
then find  the  solution for $\theta_{1}$ from \rf{sy1}.

As in Section 2,  interpreting this solution as describing a segment of a string with
 $0 < \bar \s \leq { \pi
\k\ov n}$  we may then glue
$2 n$  such pieces together  by  imposing the condition
 $\Delta \theta_0 + \epsilon \Delta \theta_1= \frac{\pi m}{n}$
 on the change of the angle.
 This  leads to   a closed
 string solution with unequal  spins. One can see that the spins  indeed differ
   by an $O(\epsilon)$ term by using the $\zeta$ coordinates (\ref{tat}) and the
  expressions for the spins  in (\ref{pqp}). Here the   two spins remain different in the large spin limit.
   Since the change in $\theta$ is small, the minimal
  energy choice for  this perturbed solution will be again  $m=1,\ n=3$.

As a result,  we find a  circular-shaped   solution
with  different spins and $\ln S$ asymptotics of the energy.
From the
leading contribution to the energy from one string arc (\ref{oan}), which is controlled by the
 asymptotic  (large $\rho$) region of the solution, we conclude that for the minimal
  solution   with $m=1,\ n=3$ we get  like in  \rf{sjk}
\begin{equation}
\mathcal{E}-\mathcal{S}= \frac{3}{2 \pi}\ln \mathcal{S}+... \ , \ \ \ \ \ \ \ \   \S= \S_1 + \S_2  \ ,   \label{oan1}
\end{equation}
but now  the subleading terms are expected to depend on $S_1 \ov S_2$.
They may be,   in principle,  determined numerically.

The  above perturbative solution thus  provides evidence
 of existence  of  a  circular-shaped   solution
with  different spins and $\ln S$ asymptotics of the energy \rf{oan1}.
To  construct it  explicitly it appears   necessary to study the general
form of the  Neumann system solution in \ci{afrt} expressed in terms of
hyperelliptic integrals.

\renewcommand{\theequation}{A.\arabic{equation}}
 \setcounter{equation}{0}

\section*{Acknowledgments }

AAT   would like to thank   A. Rej  for a question that prompted
us to revisit  2-spin  solutions in $AdS_5$. We are grateful to    L. Freyhult,  A. Irrgang,
 M. Kruczenski,
   R. Roiban, S. Zieme  and especially A. Rej
for useful discussions.

\def \bi {\bibitem}

\appendix
\subsection*{Appendix A:  Special case of the  $\kappa=\omega_2 \not=\om_1$  solution  }

The  possibility omitted in section 4.1 is when $a_{\pm}$ in \rf{koyy}  do not have the same
sign. At least one of them should be positive in order to have a periodic solution
for $x$ in \rf{koy}.
The resulting
solution will not, however,  admit  a  large spin limit.
 If $\omega
> \sqrt{2} \kappa$ and $\kappa^2 > c^2$ we get
 that $a_{+}>0$ and $-1< a_{-}<0$.
 The relations
(\ref{lbv}) are still valid.
Note that from \rf{koyy}  we have
\begin{equation}
a_{+}+a_{-}=\frac{2 \kappa^2-\om^2}{\om^2 - \kappa^2}
\end{equation}
To have  long string/large spin limit  we should
 allow  $a_{+}$ to be  large.
There are two choices of parameters  when this might be possible.
  The first one we considered already in section 4.1 --  to take $\om \approx \kappa$
  while keeping    $a_{-}$ fixed.
Since here we assume  $ 2 \kappa^2 < \om^2$, i.e. $a_+ + a_- < 0$,
we  may have $ a_+ \gg 1$ only if  $a_-$  is allowed to  take  large negative values.  However,
 from above  we have  $a_{-} > -1$, i.e.
  the maximal  value of $a_{+}$ is one, which can be reached only if $c=0$.
  The  solution discussed below   thus has a finite ``size'', not reaching the boundary.

Here   it is convenient to define the  parameters $\m,\n$ as follows
(cf. \rf{lbvv})
\begin{equation}
\m=\frac{a_{+}}{a_{+}-a_{-}}, \quad
\quad \n=\frac{a_{+}(a_{-}+1)}{a_{+}-a_{-}}, \quad \quad  a_{+}=\frac{\m-\n}{1-\m},
 \quad \quad a_{-}=\frac{\n-\m}{\m}
\end{equation}
They satisfy $0<\n,\m<1$ and $\m>\n$.
In this case  $x$   changes in the  interval:   $0 < x < \sqrt{ a_{+}}$.
Solving equation (\ref{koy}) with the initial condition $x(0)=0$ we find
\begin{equation}
x=\sinh \chi=\pm \frac{\sqrt{-a_{+} a_{-}} \sn[\frac{\om \sqrt{a_{+}-a_{-}}}{\sqrt{a_{+}
+a_{-}+2}}\sigma,\m]}{\sqrt{a_{+}-a_{-}} \dn[\frac{\om \sqrt{a_{+}-a_{-}}}{\sqrt{a_{+}
+a_{-}+2}}\sigma,\m]}
\end{equation}
Assuming again  that here $ 0 < \s \leq { \pi \ov n }$
we get
\begin{equation}
\frac{\om \sqrt{a_{+}-a_{-}}}{\sqrt{a_{+}+a_{-}+2}}=\frac{\rK[\mu] n}{\pi}
\end{equation}
Solving the equation for $\psi$  \rf{pssi}  with the initial condition $\psi(x=\sqrt{a_{+}})=0$ at
 $\sigma=\frac{\pi}{n}$ we obtain
\be
&&\psi(x)=\sqrt{\frac{a_{+}+1}{(a_{-}+1)(a_{+}-a_{-})}}\bigg[a_{-} \bigg(\Pi[\nu,\mu]-\Pi[\nu, \arcsin
 (\frac{x}{\sqrt{a_{+}}}\sqrt{\frac{a_{+}-a_{-}}{x^2-a_{-}}}),\m]\bigg) \no \\
&& \ \ \ \ \ \ \ \ \ \ \ \ +\ {\rK}[\mu] -\ {\rm F} [\arcsin (\frac{x}{\sqrt
 {a_{+}}}\sqrt{\frac{a_{+}-a_{-}}{x^2-a_{-}}}),\m]\bigg] \ ,
\ee
where
$
{\rm F}[z,\m]=\int_0^{z} \frac{d \alpha}{\sqrt{1-\m \sin^2 \alpha}}
 $  is the elliptic  ${\rm F}$-function.
At $\sigma=0$ we have  $x=0$ and
\begin{equation}
\psi(0)=\psi_0=\sqrt{\frac{a_{+}+1}{(a_{-}+1)(a_{+}-a_{-})}}\bigg(a_{-}
\Pi[\nu,\mu]+\rK[\mu]\bigg)
\end{equation}
Using (\ref{amk}) we obtain $\theta(\sigma=0)=\frac{\pi}{2}$ and $\theta(\sigma=\frac{\pi}{n})=0$.

 Thus here we get
  $\triangle \theta= \frac{\pi}{2}$. This is  different
   from the situation studied in section 4.1 since  now
the motion takes places in the interval  $0 < x < \sqrt{a_{+}}$. After mapping a segment
 in the interval $0< \sigma < \frac{\pi}{n}$ from $(\chi, \psi)$ coordinates
 into $(\rho, \theta)$  coordinates we can glue several of them
 together  to form again
  a circular  closed  string. Because
  for each arc  $\triangle \theta= \frac{\pi}{2}$ we simply have the condition
$2n =m$, where $m$ is the winding number. One can
again solve the equations $\mathcal{S}_1=\mathcal{S}_1(a_{+},a_{-},n)$, $\mathcal{S}_2=\mathcal{S}_2
(a_{+},a_{-},n)$ for $a_{\pm}$ and obtain the energy $\mathcal{E}=\mathcal{E}(\mathcal{S}_1,
\mathcal{S}_2,n)$.
We  checked numerically  that such solutions indeed exist.
The  shape of the solution for $m=1$ is illustrated in
Figure 8.
\begin{figure}
\epsfig{file=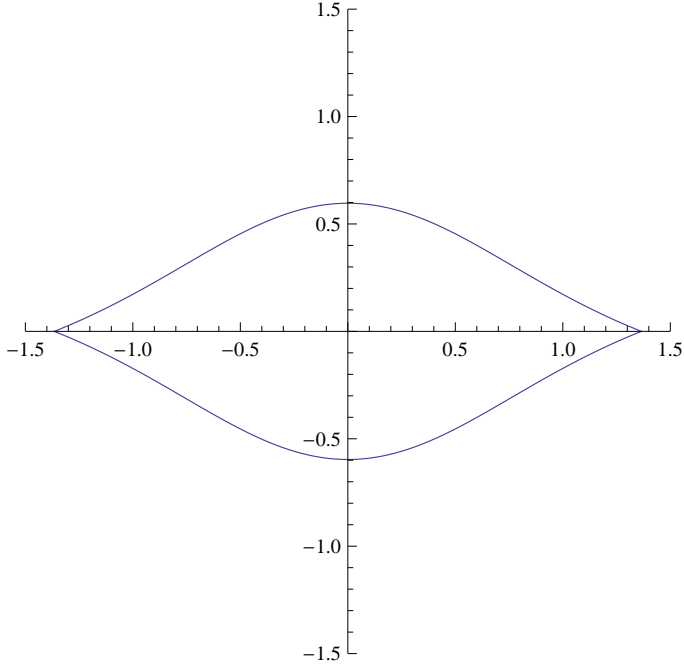, width=9cm}
\caption{$(S_1,S_2)$  string for $m=1$, $n=2$ in polar coordinates $(\rho, \theta)$
for $a_{-}=-0.5$, $a_{+}=0.4$. In  this case $\frac{S_1}{S_2}=6.614$.}
\label{fig5}
\end{figure}

Since $a_{\pm}$  change  within a  finite range
and $a_{-}$ is never close to zero, the parameters   $\om, \kappa$ cannot be large,
 and thus the spins here cannot take large values.

\renewcommand{\theequation}{B.\arabic{equation}}
 \setcounter{equation}{0}

\appendix
\subsection*{Appendix B:  Circular solution in the special case   $b_1=b_2$    }

In this Appendix we shall consider another particular case in which the solution
 can be expressed in terms of the
 elliptic integrals.
Let us briefly recall the approach of   \cite{afrt}
to the solution of  string equations for the rigid string ansatz \rf{jk}
based on the reduction  to the 1-d  Neumann integrable model.
Introducing the two  independent ``hyperbolic'' coordinates $\zeta_i$ related to $y_a$ in
 \rf{yq} by\foot{We follow
the notation in
\cite{afrt}  apart from interchanging the roles of
 $y_1$ and $y_2$  and using $\om_0$  instead of $\om_3$.
 Continuation from $S^5$ to $AdS_5$ replaces $y_3$ by $y_0$ and $\om_3 $ by $\om_0$.}
\begin{equation}
y_1^2= \frac{(\omega_1^2 - \zeta_1)(\omega_1^2-\zeta_2)}{\omega_{12}^2 \omega_{01}^2},
\ \ \ y_2^2= \frac{(\omega_2^2 - \zeta_1)
(\omega_2^2-\zeta_2)}{\omega_{12}^2 \omega_{20}^2}, \ \ \ y_0^2= \frac{(\omega_0^2 -
 \zeta_1)(\omega_0^2-\zeta_2)}{\omega_{02}^2
\omega_{01}^2} \label{tat}
\end{equation}
where $\om_a=(\om_0, \om_1,\om_2)$, $\om_0=\k$, and  $\omega_{ab}^2=\omega_a^2-\omega_b^2$,
 it was shown in \cite{afrt} that the string
equations of motion expressed in terms of  $\zeta_i(\s)$  take the form
\begin{equation}
\zeta_1'^2= -4 \frac{P(\zeta_1)}{(\zeta_2-\zeta_1)^2}, \ \ \ \ \ \ \ \ \zeta_2'^2=
-4 \frac{P(\zeta_2)}{(\zeta_2-\zeta_1)^2}  \label{ara}
\end{equation}
where
\begin{equation}
P(\zeta)=(\zeta-\omega_1^2)(\zeta-\omega_2^2)(\zeta-\omega_0^2)(\zeta-b_1)(\zeta-b_2)
\end{equation}
Here $b_1,b_2$ are two constants of motion that satisfy $b_1+b_2=\omega^2_0 + \omega_1^2+\omega_2^2
$.
They are related to  the integrals of motion of the
Neumann system $F_a$  satisfying $F_1+F_2+F_0=1$ by
\begin{equation}
b_1+b_2=(\omega_0^2+\omega_1^2)F_1+(\omega_0^2+\omega_2^2)F_2+
(\omega_1^2+\omega_2^2)F_0,  \ \ \ \
b_1 b_2=
\omega_0^2 \omega_1^2 F_1+ \omega_0^2 \omega_2^2 F_2 + \omega_1^2 \omega_2^2 F_0
\end{equation}
The integrals of motion satisfy $F_1+F_2+F_0=1$. The integrals of motion can be
expressed as
\begin{equation}\la{fff}
F_1=\frac{(b_1-\omega_2^2)(b_2-\omega_2^2)}{(\omega_0^2-\omega_2^2)(\omega_1^2-\omega_2^2)},
\ \
F_2=\frac{(b_1-\omega_1^2)(\omega_1^2-b_2)}{(\omega_0^2-\omega_1^2)(\omega_1^2-\omega_2^2)},
\ \
F_0=\frac{(b_1-\omega_0^2)(b_2-\omega_0^2)}{(\omega_0^2-\omega_1^2)(\omega_0^2-\omega_2^2)}
\end{equation}
It was argued  in \cite{afrt} that
to get  a two-spin solution  of a  circular type one needs to assume
\begin{equation}
\omega_0^2 \leq \omega_2^2 \leq \zeta_1 \leq \omega_1^2 \leq b_1 \leq \zeta_2 \leq b_2  \label{djt}
\end{equation}
while to get  a folded string solution one needs
\begin{equation}
\omega_0^2 \leq \omega_2^2 \leq \zeta_1 \leq b_1 \leq \omega_1^2  \leq \zeta_2 \leq b_2  \label{djt1}
\end{equation}
Since $P(\zeta)$ is a  degree 5 polynomial, the only way to obtain  a solution expressed in terms of
elliptic integrals
 is to assume that  at least two of the five parameters in
 (\ref{djt}),(\ref{djt1})
are equal.

 For  the circular string  choice  (\ref{djt}) the  possible special cases are:

 (i)  $\omega_1^2=\omega_2^2$  which we  considered in
 Section 2;

 (i$'$)  $\omega_1^2=\omega_2^2=\omega_0^2$ which we considered in Section 3;

 (ii) $\omega_0^2=\omega_2^2$ which we
considered in Section 4;

(iii)  $b_1=b_2$  which we shall analyze below;

 (iv) $\omega_1^2=b_1$, which reduces to the one-spin case.

\noindent
Note that the case of  $\omega_1^2=b_1=b_2$ reduces to the  one-spin case
as then  $y_1=0$.
In the folded string case (\ref{djt1})  potentially non-trivial special choices
are $\omega_2^2=b_1$ and $\omega_1^2 = b_2$; however, they   reduce to
the one-spin solution.
Note that in
  these   cases and  the case (iv)  one of the $F_a$'s in \rf{fff} is equal to zero.

Let us now
study in detail the case of $b_1=b_2\equiv b$. Then
$2 b= \omega_1^2+\omega_2^2+\omega_0^2$
 and $\zeta_2=b$. The equation (\ref{ara}) becomes
\begin{equation}
z'^2= 4 (z-\omega_2^2)(z-\omega_0^2)(\omega_1^2-z) \ , \ \ \ \ \ \ \ \ \
z \equiv \zeta_1  \ .
\end{equation}
With the initial condition $z(0)=\omega_2^2$ its  solution is
\begin{equation} \la{jpl}
z(\sigma)= \frac{\omega_2^2-p \ \omega_0^2\ \sn^2[\omega_{10}\sigma ,p]}{\dn^2 [\omega_{1
0}\sigma ,p]} , \ \ \ \  \ \ \ p\equiv \frac{\omega_{12}^2}
{\omega_{10}^2},\ \ \ \ \ \ \ 0 < p < 1
\end{equation}
We want to have a periodic motion of $z(\s)$
 between $\omega_2^2$ and $\omega_1^2$.  As in the previous
 sections,  to get a closed string solution on the interval $0 < \s \leq 2\pi$
   we  should start with \rf{jpl} on  $0 < \s \leq {\pi\ov n} $    with
   $z(0) = \om_2^2, \ \   z(\sigma=\frac{\pi}{n})=\omega_1^2$
   and  then glue together   $2n$  such segments.
  The periodicity condition implies
\begin{equation}
\om_{10}=\frac{n \rK[p]}{\pi}  \label{peri}
\end{equation}
which gives a relation between $\omega_a$'s.
The resulting expressions for the string coordinates $y_a$ in \rf{yq},\rf{tat} are
\begin{equation}
y_1^2= (\omega_1^2-z) C_1, \ \ \ \ \ \ y_2^2= (\omega_2^2-z)C_2, \ \ \ \ \ y_0^2= (\omega_0^2-z)C_0
\end{equation}
where
\begin{equation}
C_1= \frac{\omega_1^2-\omega_2^2-\omega_0^2}{2 \omega_{12}^2 \omega_{01}^2}, \ \ \ \ \ C_2=
\frac{\omega_2^2-\omega_1^2-\omega_0^2}{2
\omega_{12}^2 \omega_{20}^2}, \ \ \ \ \ C_0= \frac{\omega_0^2-\omega_1^2-\omega_2^2}{2
\omega_{02}^2 \omega_{01}^2}
\end{equation}
We need $C_1 >0$, $C_2<0$ and $C_0 <0$. This requirement along with the condition $y_0^2>1$
(cf. \rf{yq})
 gives the possible range of the frequencies
\begin{equation}
\omega_1^2-\omega_2^2 < \omega_0^2 < \omega_1^2 +\omega_2^2  \label{com}
\end{equation}
Using the relation between $y_a$ and angular   coordinates  in \rf{io} one obtains
\begin{equation}
\tan \theta= \sqrt{\frac{(\omega_2^2-z)C_2}{(\omega_1^2-z)C_1}}
\end{equation}
which implies $\theta(\sigma=0)=\frac{\pi}{2}$ and $\theta(\sigma=\frac{\pi}{n})=0$.
If $m=1,2,...$ is the winding number in  $\theta$  then
\be  n=2 m=2,4, ... \ee
For the conserved charges  (cf. \rf{enk})
\begin{equation}\la{pqp}
\mathcal{E}=\omega_0 \int_0^{2 \pi} \frac{d \sigma}{2 \pi}\ y_0^2, \ \ \ \ \  \mathcal{S}_1=\omega_1
\int_0^{2 \pi} \frac{d \sigma}{2 \pi}\ y_1^2,
\ \ \  \ \ \ \ \mathcal{S}_2=\omega_2 \int_0^{2 \pi} \frac{d \sigma}{2 \pi}\ y_2^2
\end{equation}
we then find
\begin{equation}
\mathcal{E}= - \frac{ n \omega_0 C_0 \omega_{10}}{\pi} \rE[p], \ \ \ \mathcal{S}_1=\frac{n \omega_1
 C_1 \omega_{10}}{\pi}(\rK[p]-\rE[p]), \ \ \
\mathcal{S}_2=\frac{ n \omega_2 C_2 \omega_{10}}{\pi}\big( (1-p)\rK[p]-\rE[p]\big)
\end{equation}
Given  $\mathcal{S}_1=\mathcal{S}_1(\omega_1,\omega_2,\omega_0)$,
 $\mathcal{S}_2=\mathcal{S}_2(\omega_1,\omega_2,
\omega_0)$ and the periodicity condition (\ref{peri})
we may
  solve for $\omega_a$  and  find  the energy $\mathcal{E}=\mathcal{E}(\mathcal{S}_1,
  \mathcal{S}_2,n)$.

We have checked numerically that such closed string solutions indeed exists. We illustrate
 the shape of the string with  $n=2$ in Figure 9.
\begin{figure}
\epsfig{file=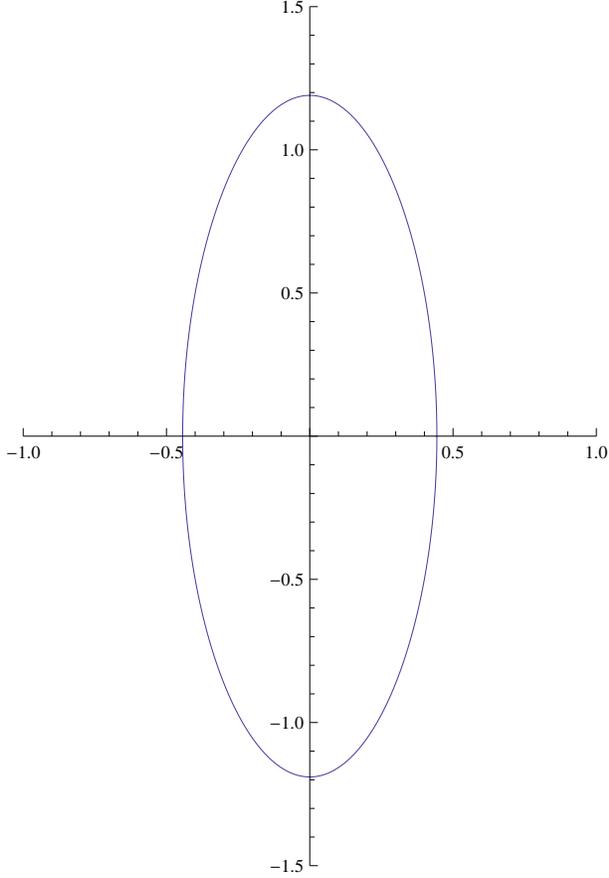, width=8cm}
\caption{$(S_1,S_2)$  string for $m=1$, $n=2$ in polar coordinates $(\rho, \theta)$
for $\omega_1=1.8$, $\omega_2=1.5$, $\omega_0=1.2867$. In  this case $\frac{S_1}{S_2}=0.144$.}
\label{fig6}
\end{figure}
In what follows we shall
 look for solutions that  admit  the possibility of
  having both spins being  large.
For finite $\omega_a$, the only limit that may give large spins is $\omega_2 \rightarrow
\omega_0$. But then  $p\rightarrow 1$ and the periodicity
condition (\ref{peri}) cannot be solved since $\rK[1]$ diverges.

The other possibility  is to take $\omega_1$ large; from (\ref{com}) this
 means that $\omega_2$ is also large, so that $p$ is small and   (\ref{peri})
 does not have  a solution for $\omega_0$.
 However, we can take  $\omega_0$
 also to be  large;  then $p$  will  be a finite number  in the interval $(0,1)$.
 Assuming
 \be \la{ott}
 \omega_1=\omega, \ \ \ \ \
  \omega_2= a \omega, \ \ \ \ \   \omega_0=b \omega, \ \ \ \ \ \
  0< b \leq  a \leq 1 \ee
  and   expanding  (\ref{peri}) in large $\omega$ we get
\begin{equation}
\omega_{10} - \frac{n \rK[p]}{\pi}= \sqrt{1-b^2} \omega- \frac{n}{\pi}\rK[\frac{a^2-1}{b^2-1}]+
O(\frac{1}{\omega}) =0
\end{equation}
which leads, to leading order in  $\om\gg 1 $,  to $a=b=1$.
This suggests to set\foot{Including $O(\om^0)$ terms in $\om_i$   does not lead to new
solution.}
\begin{equation}\omega_0=\omega- \frac{c}{\omega} + ..., \ \ \ \
\omega_2=\omega- \frac{d}{\omega}+ ..., \ \ \ \ \omega_1=\omega \gg 1 \ .
\end{equation}
Then the periodicity condition (\ref{peri}) gives
\begin{equation}
\sqrt{2 c}- \frac{n}{\pi}\rK[\frac{d}{c}]+O(\frac{1}{\omega})=0  \label{peri1}
\end{equation}
Let us define  $q=\frac{d}{c} < 1 $, so that
 $p=q+O(\frac{1}{\omega^2})$.
Expanding the spins  and the energy  in large $\omega$ gives
\begin{equation}
\mathcal{S}_1= \frac{n \omega^3}{4 \pi \sqrt{2 c}}\frac{ \rE[q]-(1-q)\rK[q]}{d (1-q)}+O(\omega), \ \ \ \ \ \
\mathcal{S}_2= \frac{n \omega^3}{4 \pi \sqrt{2 c}}\frac{
\rK[q]-\rE[q]}{d}+O(\omega)  \label{spp1}
\end{equation}
\begin{equation}
\mathcal{E}= \frac{n \omega^3}{4 \pi \sqrt{2 c}}\frac{\rE[q]}{c(1-q)}+O(\omega)
\end{equation}
For this solution  both spins are thus  large
(scaling as $\om^{1/3}$) and different
with (cf. (\ref{peri1}))
\begin{equation}
\mathcal{E}-\mathcal{S}_1-\mathcal{S}_2= \frac{3 n \omega}{4 \pi \sqrt{2c}}\rK[q]+O(\frac{1}{\omega})=
 \frac{3}{4}\omega + O(\frac{1}{\omega}) \ ,
\end{equation}
\begin{equation}
\frac{\mathcal{S}_1}{\mathcal{S}_2}=\frac{\rE[q]-(1-q)\rK[q]}{(1-q)(\rK[q]-\rE[q])}
+ O(  \frac{1}{ \om^2}) \label{qas}
\end{equation}
We have checked that if  $\frac{\mathcal{S}_1}{\mathcal{S}_2} \geq 1$   the equation  (\ref{qas})
 has a solution   for  $q$. Then we can use (\ref{peri1})
 to find
  $c$.
 At large $\omega$  we have
\begin{equation}
\omega^3= 4 c(1-q)\frac{\rK[q]}{\rE[q]} \mathcal{S}, \ \ \ \ \ \ \ \ \ \  \mathcal{S}\equiv
\mathcal{S}_1+\mathcal{S}_2
\end{equation}
which leads to the following  dependence of the energy on the total spin
\begin{equation}
\mathcal{E}-\mathcal{S}= \frac{3}{4}\bigg(4c (1-q)\frac{\rK[q]}{\rE[q]}\bigg)^{1/3} \mathcal{S}^{1/3}+... \ , \
\ \ \ \ \ \ \ \S \gg 1  \  ,
\end{equation}
where $c,q$ are functions of  $\S_1 \ov \S_2$ and $n$.

The particular case $\S_1=\S_2$
  corresponds to  $q=0$,   i.e.   $d=0$ and $c=\frac{n^2}{8} = { m^2 \ov 2} $.
Then
\begin{equation}
\mathcal{E}-\mathcal{S}=\frac{3}{4}\big( 2m^2\big)^{1/3} \mathcal{S}^{1/3} + ...
\end{equation}
i.e.  we  recover the round  circular string \ci{ft2} expression
 \rf{faa}  with the winding number
 $m={ n \ov 2} = 1,2,..$.

\def \bi {\bibitem}


\end{document}